\documentclass[reprint, amssymb, amsmath, aip, jcp]{revtex4-1}
\usepackage{multirow}
\usepackage{graphicx}%
\usepackage{docs}%
\usepackage{bm}%
\usepackage[colorlinks=true,linkcolor=blue]{hyperref}%

\begin{document}

\author{Madhusmita Tripathy}
\affiliation{Department of Physics, Indian Institute of Technology Madras, India}
\author{P. B. Sunil Kumar}
\affiliation{Department of Physics, Indian Institute of Technology Madras, India}
\author{Abhijit P. Deshpande}
\affiliation{Department of Chemical Engineering, Indian Institute of Technology Madras, India}
\email{abhijit@iitm.ac.in}


\title{Molecular Structuring and Percolation Transition in Hydrated Sulfonated Poly(ether ether ketone) Membranes}


\begin{abstract}

The extent of phase separation and water percolation in sulfonated membranes are the key to their performance in fuel cells. Toward this, the effect of hydration on the morphology and transport characteristics of sulfonated poly(ether ether ketone), sPEEK, membrane is investigated using atomistic molecular dynamics (MD) simulation at various hydration levels ($\lambda$: number of water molecules per sulfonate group) between 4 and 15. At the molecular level, the evolution of local morphology is investigated in terms of structural pair correlations and minimum pair distances, and the transport properties are studied in terms of mean squared displacements (MSDs) and diffusion coefficients. The water-sulfur interaction in sPEEK is found to be stronger than that in Nafion, as observed in experiments. As opposed to Nafion, a weaker interaction of  hydronium, with sulfonate, than water is observed. The behavior of water in sPEEK membrane is found to remain far from bulk as indicated by its diffusion coefficient.  Analysis of simulation data indicate that at low $\lambda$, the largest water cluster forms a narrow connected path of water molecules and hydronium ions. With increasing $\lambda$, larger water domains appear, spanning more than half of the simulation box at $\lambda$ = 15.  Small isolated clusters are present at all hydration levels, demonstrating the extent of phase separation in sPEEK to be lesser than that in Nafion. Various analyses, both at molecular and collective level, suggest the occurrence of a percolation transition between $\lambda$ = 8 and 10, which leads to a connected network of water channels in the membrane, thereby boosting the mobility of hydronium ions. 

\end{abstract}

\maketitle
\section{Introduction}

Polymer electrolyte membrane (PEM) fuel cell has shown promising potential in the ongoing quest for reliable alternate energy sources. The main component of such a fuel cell is a polymer membrane separating the anode and cathode, which allows only protons to pass through it, while blocking the fuel from crossing over. Sulfonated poly(ether ether ketone), sPEEK,  is  an emerging alternative to the widely used Nafion \citep{Mauritz} as a material for PEM. sPEEK-based PEM  has been reported to possess high thermal stability \citep{speek_thermal_stability} and proton conductivity comparable to that based on Nafion \citep{high_temp_speek}, along with narrower and less connected hydrophilic channels that can be useful in reducing fuel cross-over \citep{Kreuer}.  At sufficiently hydrated state, these  sulfonated polymer membranes attain a phase separated morphology, which gives rise to connected network of water channels for proton conduction  \citep{Mauritz, Kreuer}.

While the origin and morphology of the water channels in the sulfonated polymer membranes still remain under investigation, scattering experiments have  predicted them to be of the order of few nano-meters (nm) in diameter \citep{Kreuer,  Mauritz, Schmidt-Rohr}. Simulations aiming to probe the membrane at such large length scales  usually rely on mesoscale particle-based methods such as dissipative particle dynamics (DPD) or mean-field theory-based dynamic density functional theory (DDFT).  Such  studies have been extensively used to understand the morphology  of Nafion  \citep{DPD_amphiphilic, DPD_nafion, Wescott_Nafion} and  sPEEK  \citep{DPD_speek, Komarov_1, Komarov_2} membranes. These  simulations were quite successful in  providing  the  mesoscale morphology of the sulfonated membrane, in terms of the phase separation and its evolution with increasing water content.  However, the  structural changes happening at the molecular scale,  which is the key reason for the  phase separated morphology, cannot be captured by these simulations, owing to their intrinsic length scales that are relatively larger.

On the other hand, atomistic simulations are helpful in the study of the dynamics of water and counter-ion  and in understanding the membrane morphology at the level of individual sulfonate groups, counter-ions, and water molecules \citep{Mahajan_1, Mahajan_2, Devnathan_nafion_1, Devnathan_nafion_2, Devnathan_nafion_3, Devnathan_phspeekk, Brunello_1, Brunello_2, Bahlakeh2012}. Such studies can also be used to understand the effect of hydration, temperature, and degree of sulfonation on the static and dynamical properties of the hydrated  membranes.  However, computational cost limits these simulations to small system sizes. In an atomistic simulation on Nafion and Hyflon using a simulation box size of $\sim$ 7.5 nm, Karo et al. \citep{Karo_large_system} reported the dependency of water channel topology and connectivity on the simulated system size, by comparing them against those obtained in their earlier work with a smaller system size (nearly half of the larger box size) \citep{Karo_small_system}. Though achieving larger length scales in atomistic simulation is nonetheless possible, it is computationally expensive and calls for sophisticated modeling approach and computing resource \citep{Voth_large_aa, Komarov2013_large_aa}. In the work presented here, we use a systematic coarse-graining method developed by us \citep{cg_speek_madhu} to access the atomistic configurations with  large system sizes.

In accordance with the scattering experiments \citep{Kreuer, Mauritz}, simulations on sulfonated membranes have reported the presence of small water domains at low hydration level $\lambda$, the number of water molecules per sulfonate group. These domains are believed to grow with increasing $\lambda$ and eventually get connected to form a percolating network of  water channels at $\lambda_p$. Devanathan et al. have used atomistic simulation to characterize water percolation in Nafion and phenylated sPEEKK membranes. For fully sulfonated Nafion, they have reported a $\lambda_p$ value between $\lambda$ = 5 and 6 \citep{Devnathan_nafion_1, Devnathan_nafion_3} for 100\% sulfonated chains with total 40 SO$_3^-$ units at 300 K. For fully sulfonated phenylated sPEEKK in presence of 1.33 M methanol with 180 SO$_3^-$ units at 360 K, they reported a $\lambda_p$ value of 7.92 \citep{Devnathan_phspeekk}.

System size, degree of sulfonation and dissociation, temperature, and presence of co-solvent such as methanol, are all factors which can affect the percolation transition in the membrane. Using density functional theory, Komarov et al. \citep{Komarov_2} have  reported a $\lambda_p$ value close to 9 for sPEEK membrane with 50\% sulfonation at 300 K. Though there exist several atomistic simulation studies on the effect of hydration, to the best of our knowledge, there has been no such systematic studies on water percolation in sPEEK. Brunello et al. \citep{Brunello_1} used atomistic simulations, on a 40\% sulfnated sPEEK system with 80 SO$_3^-$ units at 353.15 K, to study the hydration effect in terms of radial distribution functions (RDFs), coordination numbers, and hydronium diffusivity.  Mahajan and Ganesan \citep{Mahajan_1} also studied the effect of methanol, using a  50\% sulfonated sPEEK system with 92 SO$_3^-$ units in presence of 1 M methanol at 338 K.  A   comparison between the different studies is very important for a better characterization and understanding of the effect of hydration on the membrane morphology.

The present study intends to systematically characterize the extent of phase separation and percolation in sPEEK membrane, with increasing water content. At the level of individual sulfonate group, water molecule, and hydronium ion, the evolution in their local arrangement is studied in terms of various RDFs and the minimum pair distances.  The transport properties of water molecule and hydronium ion are studied in terms of their mean-squared displacements (MSDs) and diffusion coefficients. Detailed cluster analysis is performed to characterize the collective water phase in the membrane. The variations in these properties with increasing hydration is  analyzed to characterize the morphological evolution.

Our main results are the following. The water molecules are found to strongly interact with the sulfonate groups in the sPEEK membrane, in agreement with experimental prediction. The hydronium ions exhibit a comparatively weaker interaction with the sulfonate groups, as opposed to that in Nafion.  The water phase, though grows with increasing hydration, remains distinct from bulk phase, as indicated by the diffusion coefficient. A percolation transition is identified in the sPEEK membrane between $\lambda$ = 8 and 10, which leads to enhanced mobility of hydronium ion. However, small water domains remain isolated in sPEEK membrane even at a large water content, as opposed to that in Nafion, demonstrating the extent of phase separation to be lesser in the former than Nafion. The rest of the paper is organized as follows. In Section 2 , we describe the details of the atomistic simulation. In Section 3, we discuss the results obtained from analyses on sPEEK membrane at various water contents. We conclude in Section 4 with a short summary and probable future directions.

\section{MODELING AND SIMULATION}

\begin{figure}
\centering
 \includegraphics[scale =0.2]{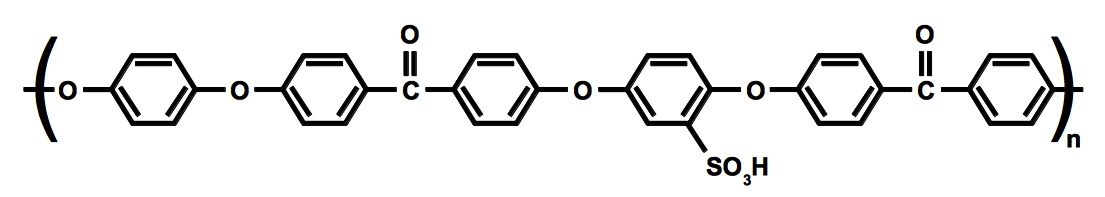}
  \caption{Schematic of a sPEEK repeat unit with alternate sulfonated units.}
  \label{fgr:structure}
\end{figure}

In our earlier work on systematic coarse-graining of sPEEK \citep{cg_speek_madhu}, we developed a model, with the highest possible coarse-grained (CG) mapping, for a 50\% sulfonated sPEEK with alternate sulfonated units (Figure~\ref{fgr:structure}).

\begin{figure}
  \includegraphics[scale= .3]{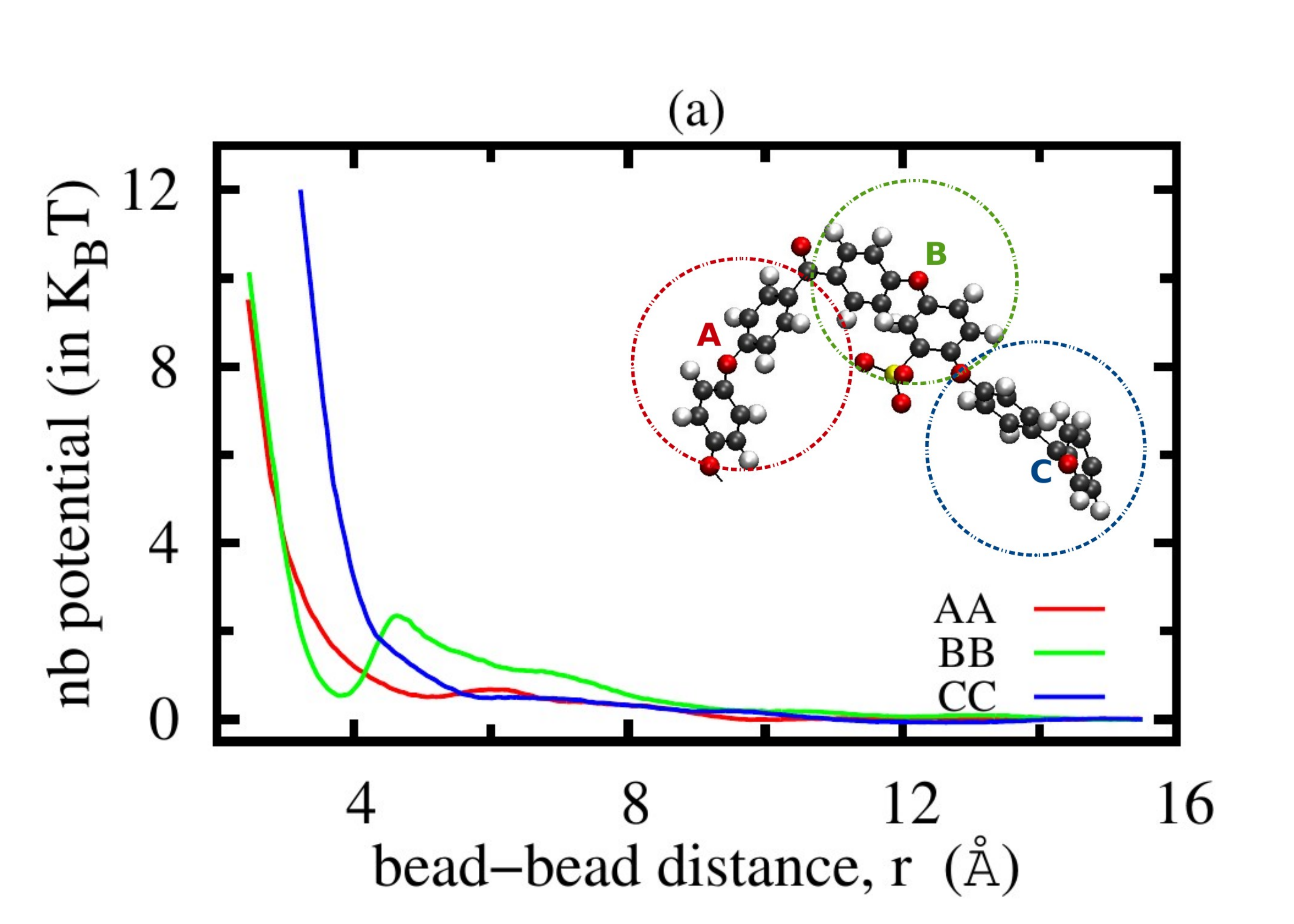}
  \includegraphics[scale= .3]{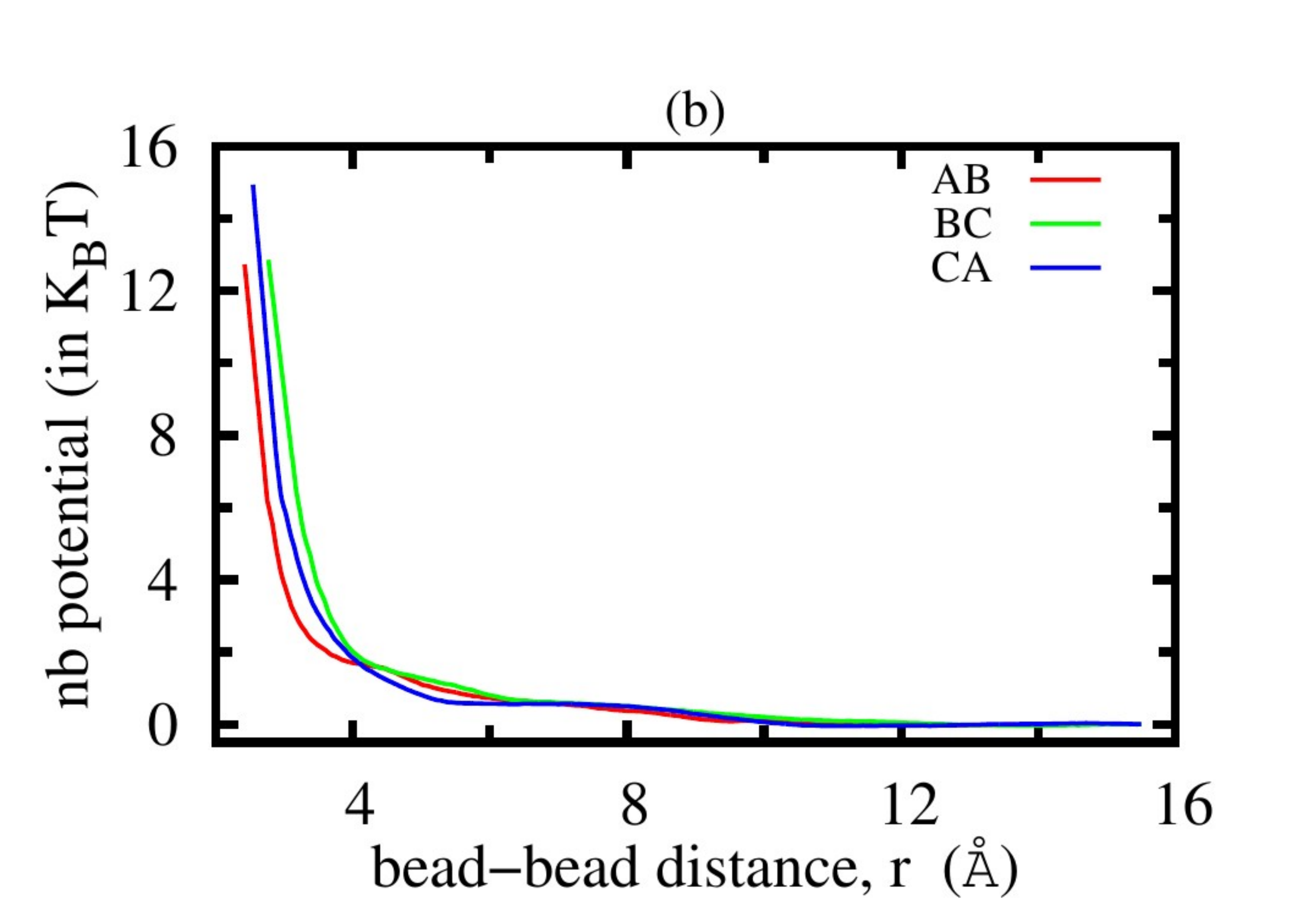}
  \caption{The optimized non-bonded (nb) potentials between  (a) similar, and  (b) dissimilar CG bead pairs. Inset in (a) shows the CG mapping of the atomistic repeat unit, shown in Figure~\ref{fgr:structure}, into 3 CG beads: A, B, and C.} 
   \label{fgr:cg_map_pot_nb}
\end{figure}

Our CG model coarse-grains the atomistic repeat unit shown in Figure~\ref{fgr:structure}, consisting of 71 atoms, into 3 CG beads, A, B, and C, as shown in the inset of Figure~\ref{fgr:cg_map_pot_nb}(a). The corresponding CG potentials were optimized using iterative Boltzmann inversion \citep{IBI} method. The optimized non-bonded (nb) potentials between various CG bead pairs are shown in Figure~\ref{fgr:cg_map_pot_nb}. These potentials along with bond, angle, and dihedral potentials, constitute the full set of numerical potential for the sPEEK system under consideration. Though obtained from base atomistic simulation of short single chain, the CG potentials were shown to be scalable to larger systems,  both for  single chains of longer length as well as for multiple chains, at the same environmental conditions. Along with accelerated equilibration of comparatively larger CG systems, CG simulation can reproduce the target distributions collected from the atomistic simulation. The corresponding equilibrated atomistic trajectory was easily obtained upon back-mapping, wherein a pre-equilibrated atomistic chain is mapped onto an equilibrated CG configuration of a comparatively larger system size, followed by short Molecular Dynamics (MD) run. Further details on the coarse-graining methodology, CG potentials, and back-mapping can be obtained in the original reference \citep{cg_speek_madhu}.

Using the above mentioned protocol, we back-mapped equilibrated CG structure consisting of 8 sPEEK chains, each with 30 PEEK-sPEEK repeat units (50\% degree of sulfonation).  240 hydronium ions were randomly introduced in the identified voids  for charge neutrality, along with 2400 water molecules, at hydration level $\lambda$ = 10 (the number of water molecules per sulfonate group), resulting in the initial configuration for further atomistic simulation. This method can thus be useful in generating near-equilibrium atomistic configuration of larger system size, without the need of amorphous builder tools, which are generally used in such simulation studies \citep{Mahajan_1, Mahajan_2, Brunello_1, Brunello_2}. The interaction parameters for the polymer were taken from Dreiding force field \citep{Dreiding}. For water molecules, the modified TIP3P model \citep{TIP3P} was used, while the force field from Jang et al. \citep{Jang} was adopted for hydronium ions. MD simulation was performed using LAMMPS (large-scale atomic/molecular massively parallel simulator) package \citep{Lammps} and a Verlet type integrator with integration time step of 1 femto-second (fs). The state point was fixed at 400 K temperature and 1 atm pressure using Nos\'e-Hoover thermostat and barostat, with time constants of 0.1 pico-second (ps) and 1 ps, respectively. To relax the stretched/compressed bonds and overlapping polymer segments, and to accelerate equilibration, repeated thermal annealing cycles were performed. The simulation run lasted for 11.2 nano-seconds (ns) with snapshots of trajectory collected at every 10 ps.  Stable energy and volume trajectories were recorded after 5 ns run, and the equilibrium density was calculated to be 1.203 gm cm$^{-3}$.

The CG potentials for sPEEK system were originally extracted using a single chain atomistic system at  $\lambda$ = 4.067  \citep{cg_speek_madhu}.  However, we show that the back-mapped structure can be used as the starting configuration for (larger) atomistic simulations at other hydration levels as well. As a test case, we performed 20 statistically independent atomistic simulations on single chain sPEEK systems, with 30 hydronium ions at $\lambda$ = 10. The average equilibrium density and RDFs, between the sulfur atoms, water molecules, and hydronium ions, were computed for both the multi-chain and single chain systems for comparison.  The density was found to be 1.168 $\pm$ 0.006 gm cm$^{-3}$, close to that from the multi-chain counter-part. The overall agreement, between the multi-chain RDFs and the averaged (over 20 systems) single chain ones, was found to be quite good (data not shown), even though the comparison was against a set of averaged distributions calculated from statistically independent systems. This suggests a general scheme for simulating comparatively larger atomistic systems possessing well-mixed polymer configuration, starting from an equilibrated CG structure.  Also, such a scheme does not call for much complicated equilibration techniques or long simulation runs.

To understand the morphological changes happening in the sPEEK membrane due to  variation in water content, we simulated  multi-chain atomistic systems of sPEEK at hydration levels $\lambda$ = 4.067($\sim$4), 7, 8, 9, 12, and 15, at 400 K temperature and 1 atm pressure, using the above mentioned methodology. The above temperature was chosen as sPEEK has been considered as an alternative to Nafion, especially for high temperature performance \citep{high_t_pem}. While excellent ionic conductivity for sPEEK membranes at 393 K has been reported in experiment \citep{high_temp_speek}, the previous simulation studies on sPEEK have investigated the membrane properties only up to 353 K \citep{Mahajan_1, Brunello_2}.  We chose to simulate hydrated sPEEK membranes at an elevated temperature,  of 400 K, as  such an elevated temperature also quickens the equilibration in atomistic simulation. 

The equilibrium box sizes, density values, and other important details of the simulation are summarized in Table~\ref{tbl:multi_chain_sPEEK_aa}. The equilibrium volume of the simulation box was found to monotonously increase with increasing hydration. The density, however, showed an initial increase followed by steady decrease, possibly due to a combination of molecular re-organization and swelling in the membrane \citep{swelling_saxs}. Experiments suggest the density of dry PEEK to be in the range of 1.26-1.32 gm cm$^{-3}$ \citep{Huang} and the density is known to decrease upon sulfonation  \citep{Zaidi} and  hydration \citep{Mahajan_1}. The equilibrium density values, reported here, are therefore well within the expected range for hydrated sPEEK. Brunello et al. \citep{Brunello_1} have reported  sPEEK membrane densities at 353.15 K, ranging from 1.07 gm cm$^{-3}$ at $\lambda$ = 4.9 to 1.18  gm cm$^{-3}$ for $\lambda$ = 11.1. sPEEK membrane densities, in presence of 1 M methanol at 338 K, has been reported by Mahajan and Ganesan \citep{Mahajan_1} to  range between 1.1318 gm cm$^{-3}$ at  $\lambda$ = 1.94 to 0.8587 gm cm$^{-3}$ at  $\lambda$ = 25.05. Various results obtained from the detailed analysis of these simulations are presented in the next section.

\begin{widetext}
\begin{center}
\begin{table}
  \caption{Simulation details of multi-chain atomistic systems of sPEEK.}
  \label{tbl:multi_chain_sPEEK_aa}
\begin{tabular}{c c c c c}
\hline \hline
Hydration & Number of water & Total number & Equilibrium & Equilibrium  density\\ 
level ($\lambda$) & molecules (N$_\text{W}$)  & of atoms & box size (nm) & (gm cm$^{-3}$)\\
\hline
4.067 & 976 & 20944 & 6.315 &1.183 \\
7 & 1680 & 23056 & 6.433 & 1.199 \\
8 & 1920 & 23776 & 6.432 & 1.226 \\
9 & 2160 & 24496 & 6.503 & 1.212 \\
10 & 2400 & 25216 & 6.566 & 1.203 \\
12 & 2880 & 26656 & 6.683 & 1.189 \\
15 & 3600 & 28816 & 6.840 & 1.177 \\
\hline
\hline
\multicolumn{3}{l}{Number of polymer chains} & \multicolumn{2}{c}{8} \\
\multicolumn{3}{l}{Degree of polymerization} & \multicolumn{2}{c}{60} \\
\multicolumn{3}{l}{Degree of sulfonation} & \multicolumn{2}{c}{50} \\
\multicolumn{3}{l}{Number of sulfonate groups (N$_\text{S}$)} & \multicolumn{2}{c}{240} \\
\multicolumn{3}{l}{Number of hydronium ions (N$_\text{H}$)} & \multicolumn{2}{c}{240} \\
\multicolumn{3}{l}{Temperature} & \multicolumn{2}{c}{400 K} \\
\multicolumn{3}{l}{Pressure} & \multicolumn{2}{c}{1 atm} \\
\hline  \hline
\end{tabular} 
\end{table}
\end{center}
\end{widetext}
\section{RESULTS AND DISCUSSION}

In this section, we present the important results obtained from the atomistic simulations on sPEEK membrane at various hydration levels. The effect of hydration on membrane morphology and phase separation is investigated in terms of local structuring of individual atomistic units.  The effect on the transport properties of water and hydronium ions is discussed in terms of their MSDs and diffusion coefficients. Finally, the growth of water phase is analyzed in terms of clustering of water molecules and hydronium ions.

\subsection{Effect of hydration on the structuring of sulfonate groups, water molecules, and hydronium ions }

The effect of hydration on the local structuring of sulfonate groups, water molecules, and hydronium ions was studied in terms of various RDFs. As the changes in RDFs can sometimes be subtle, we also analyzed the averaged minimum distances between a pair of species, to provide further insights into hydration effects.


\subsubsection{Radial distribution functions}

In Figure~\ref{fgr:swh_rdf}, we plot the RDFs, calculated between the sulfur atoms, and the oxygen atoms of the water molecules and hydronium ions for various hydration levels. To  highlight the variation, the RDF is  multiplied by the mean number density of the participating species ($\rho_\text{W}$/$\rho_\text{h}$/$\rho_\text{S}$; for water, hydronium, and sulfur, respectively) at the corresponding hydration level.  

The peaks in sulfur-water (SW) RDF, as shown in Figure~\ref{fgr:swh_rdf}(a), became sharper with increasing $\lambda$, indicating water to get more and more structured around the sulfonate groups. The first  peak, positioned around 3.65 \AA, was followed by a weak shoulder around 4.2 \AA, which became prominent as $\lambda$ increased. Such a feature in SW RDF for sPEEK, can possibly arise due to the various possible orientations of water molecules that take part in hydrogen bonding with the sulfonate group, as previously studied for sulfate anion (SO$_4^{2-}$) \citep{sulfate_anion}.  A similar feature in SW RDF was also observed by Mahajan and Ganesan \citep{Mahajan_1}.

While the position of the first peak in sulfur-hydronium (SH) RDF (Figure~\ref{fgr:swh_rdf}(b)) was found to be unaffected, its height decreased with increasing $\lambda$. On the other hand, the broad second peak did not posses any appreciable change with $\lambda$. Thus, change in water content is found to strongly affect the hydronium ions which lie within the first shell around the sulfonate group, while those in the second shell remain mostly unaffected.

\begin{widetext}
 \begin{center}

\begin{figure}
  \includegraphics[scale= .64]{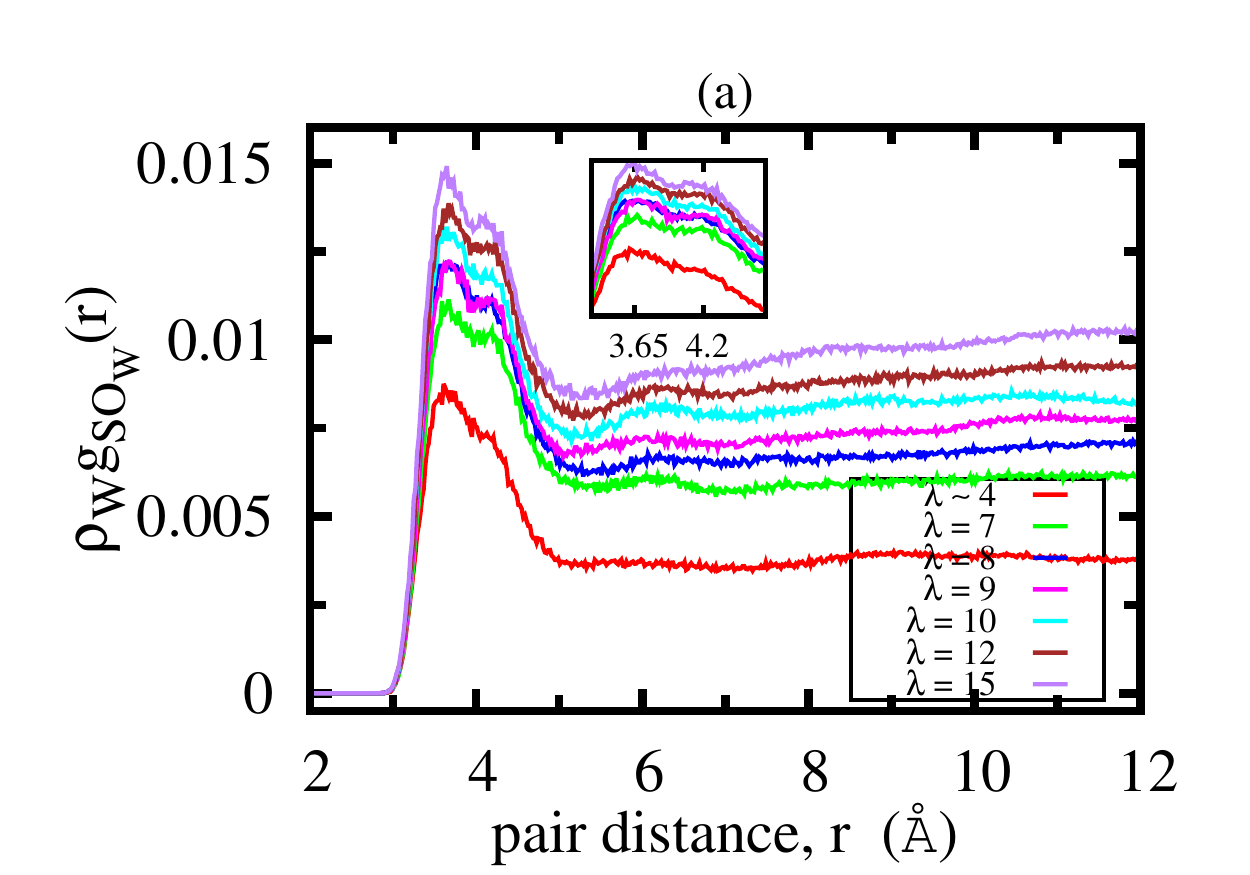}
    \includegraphics[scale= .64]{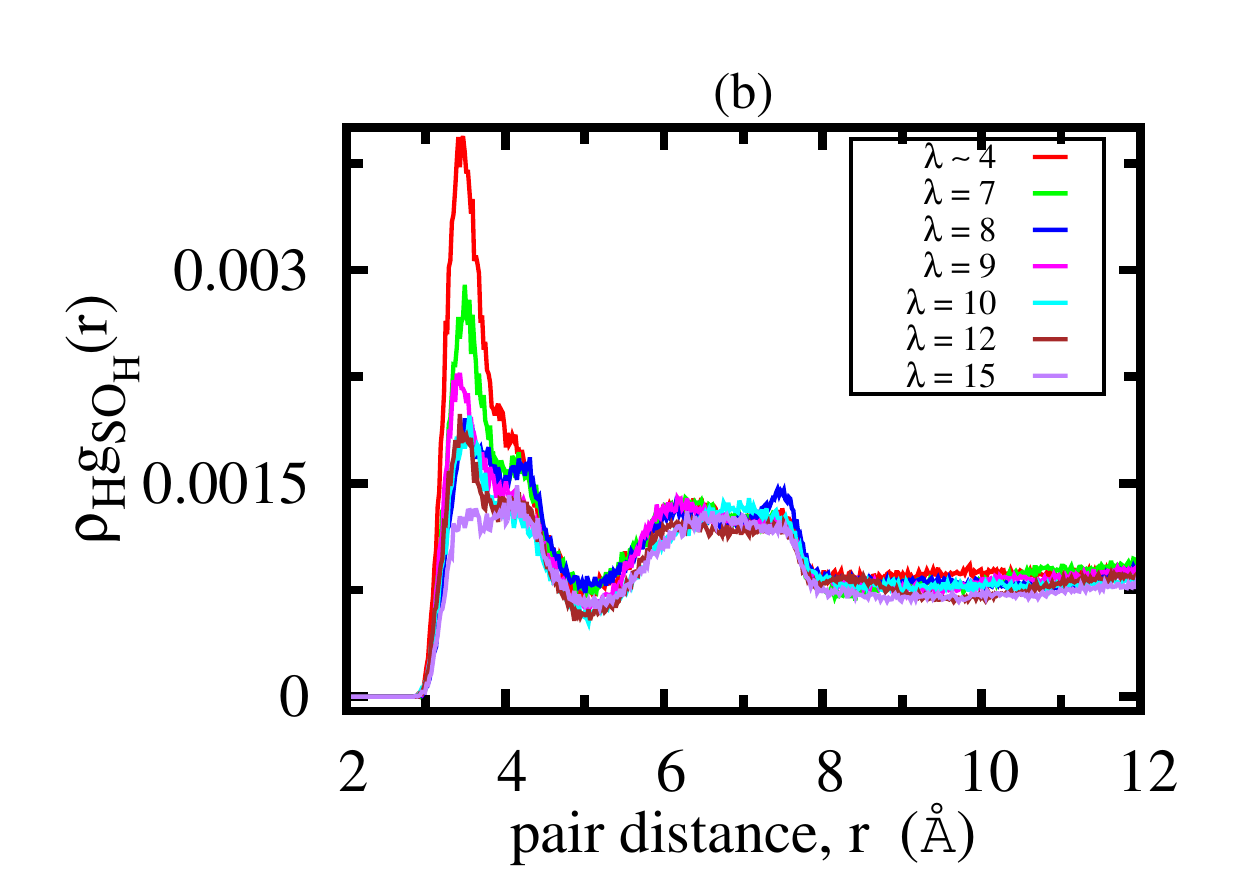}
    
  \includegraphics[scale= .64]{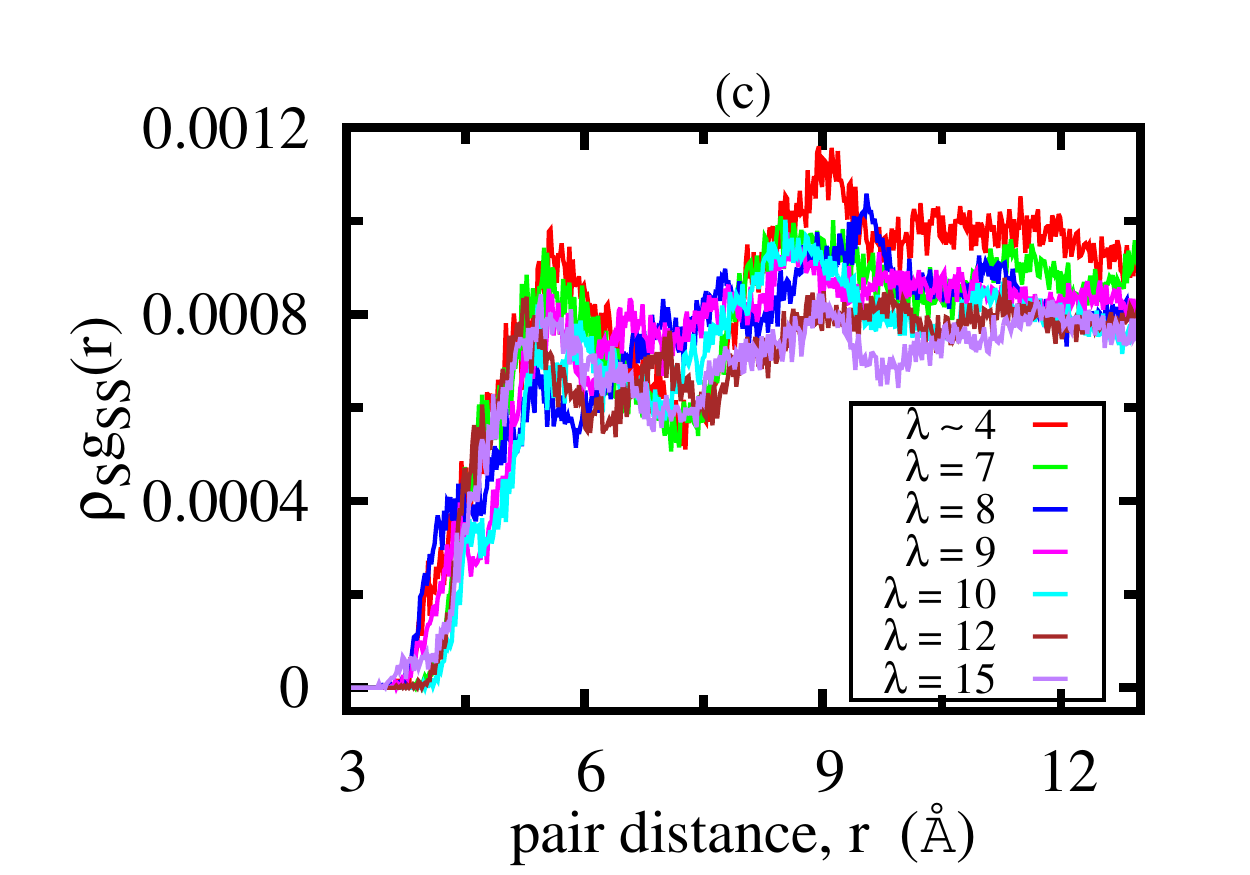}
  \includegraphics[scale= .64]{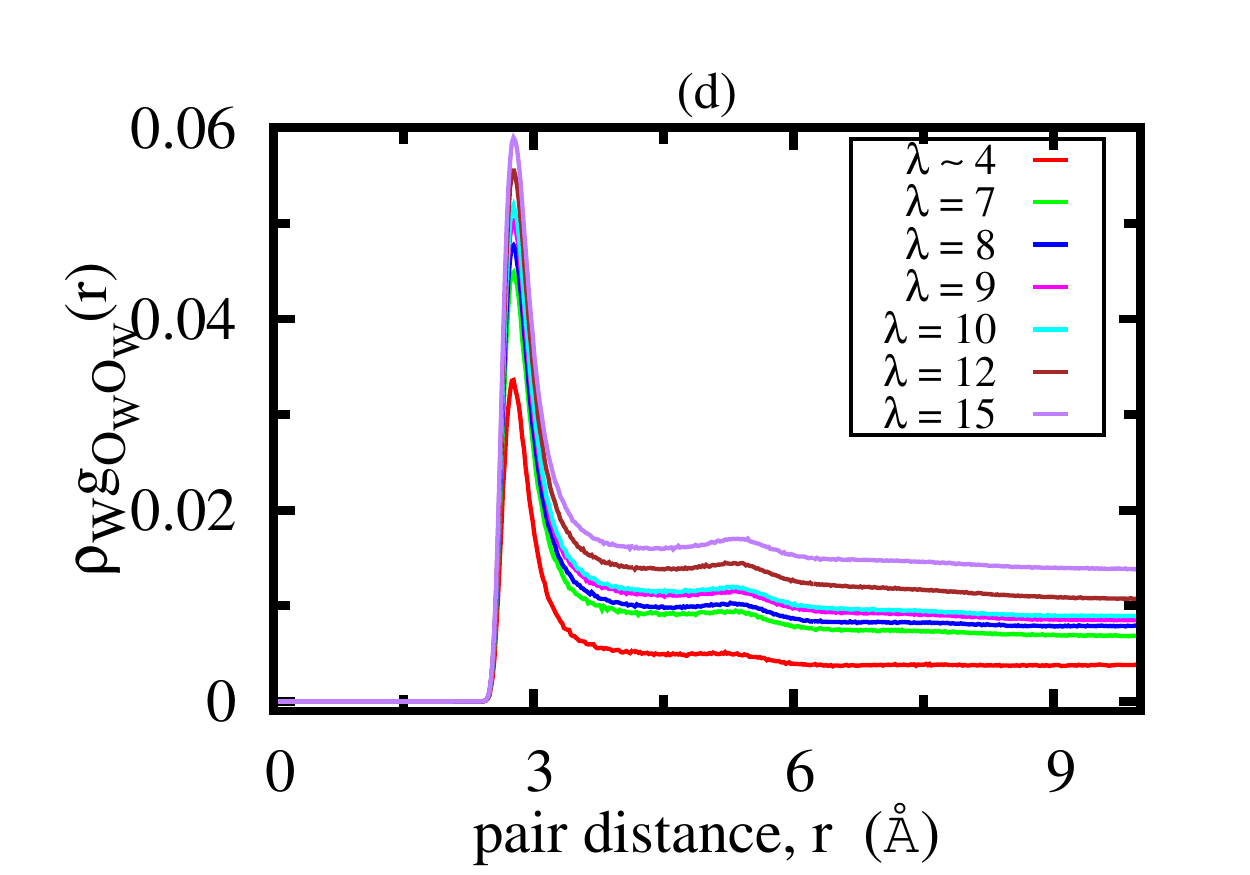}
 
 \includegraphics[scale= .64]{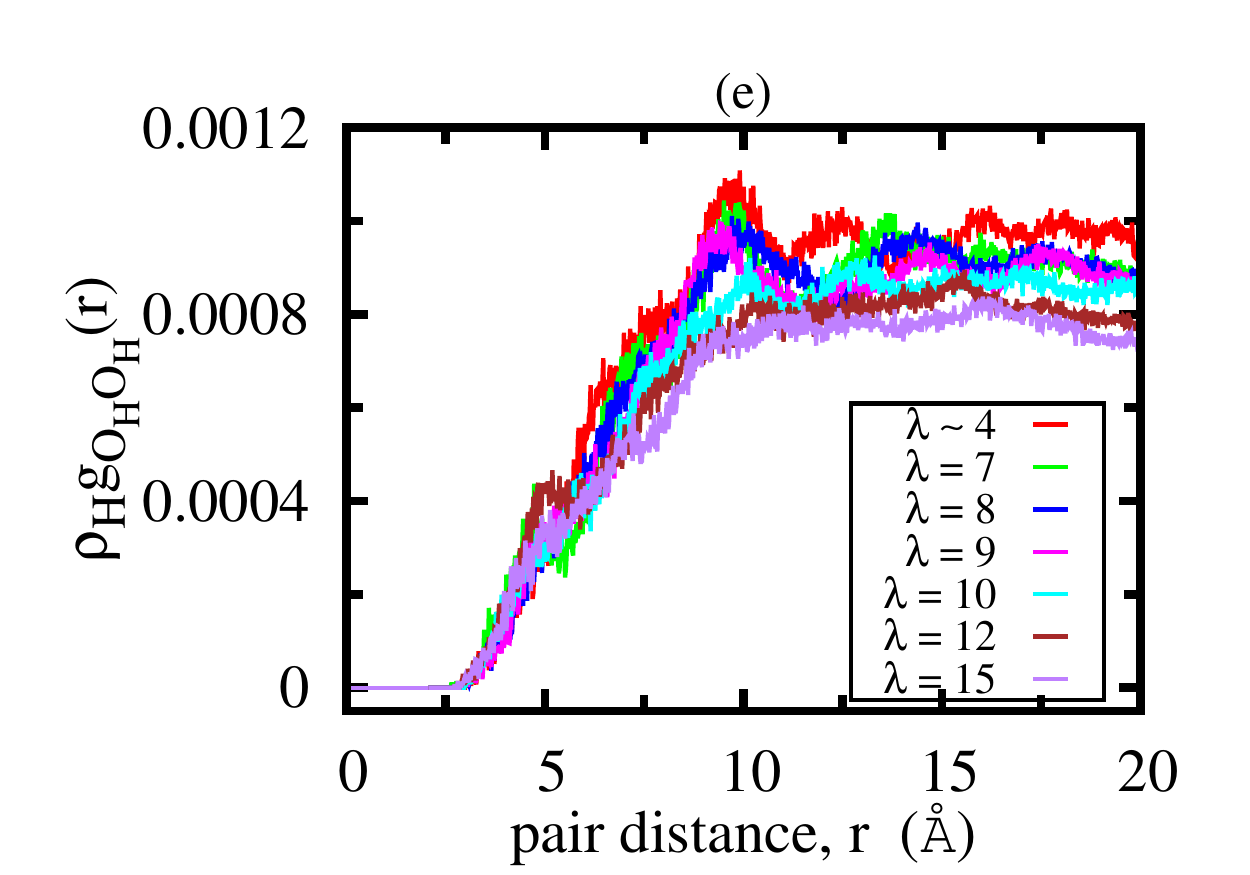}
  \includegraphics[scale= .64]{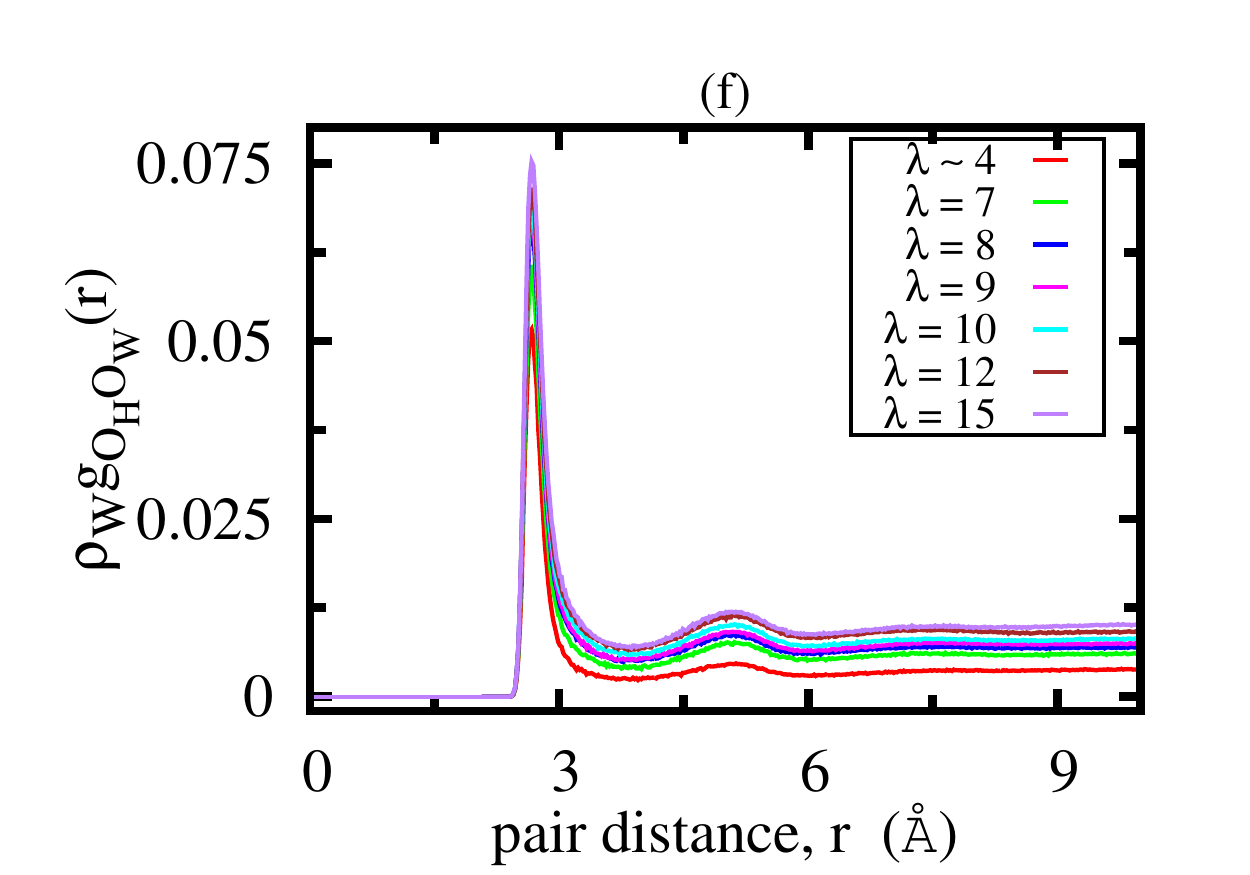}
  \caption{(a)sulfur-water (SW), (b)sulfur-hydronium (SH), (c)sulfur-sulfur (SS), (d)water-water (WW), (e) hydronium-hydronium (HH), and (f) hydronium-water (HW) RDF for all hydration levels under study at 400 K. In all case, the RDF is  multiplied by the mean number density of the participating species ($\rho$) at the corresponding hydration level. For water molecules and hydronium ions, the oxygen atoms were considered as the reference site.}
   \label{fgr:swh_rdf}
\end{figure}
 \end{center}

\end{widetext}

The first peak in sulfur-sulfur (SS) RDF (Figure~\ref{fgr:swh_rdf}(c)), positioned around 5.5 \AA, also did not shift appreciably with hydration. This is due to the rigid nature of the aromatic polymer backbone to which the sulfonate groups are attached and is in accordance with earlier findings on sPEEK-based membranes \citep{Mahajan_1, Devnathan_phspeekk}. Though the height of the first peak did not change much, that of the second peak (positioned around 9 \AA) decreased with increasing $\lambda$, leading to comparable peak heights at $\lambda$ = 15, and indicating the depletion of neighboring sulfur atoms in the second shell. The solvation of the sulfonate groups in sPEEK with increasing hydration (as indicated by SW RDF) leads to subsequent sparseness of the sulfonate groups, which move farther away from each other as $\lambda$ increases.

The peaks in water-water (WW) RDF (Figure~\ref{fgr:swh_rdf}(d)) were found to increase in height as $\lambda$ increased,  indicating strong structuring of water with increasing hydration. The first peak was positioned close to the experimental oxygen-oxygen distance for a water dimer (2.73 \AA) \citep{TIP3P}. While the overall structure in hydronium-hydronium (HH) RDF (Figure~\ref{fgr:swh_rdf}(e)) became featureless,  the peaks in hydronium-water (HW) RDF (Figure~\ref{fgr:swh_rdf}(f)) increased in height as the water content in the membrane increased. These observations suggest the water organization to become more structured with increasing $\lambda$, leading to development of a collective water phase in the membrane. The hydronium ions loose their original structuring and the hydronium-water interaction becomes stronger with increasing water content. This is again in agreement with the observations from SW and SH RDFs, which suggested the sulfur-hydronium interaction to become weak and the hydronium ions to get depleted into the growing water phase with increasing $\lambda$.

\subsubsection{Averaged minimum pair distances}

For a given equilibrated atomistic configuration, the average distance between a sulfonate group (centered on the sulfur atom) and its nearest water molecule or hydronium ion (centered at the oxygen atom) is calculated by averaging over all such pairs. 

\begin{figure} 
  \includegraphics[scale= .64]{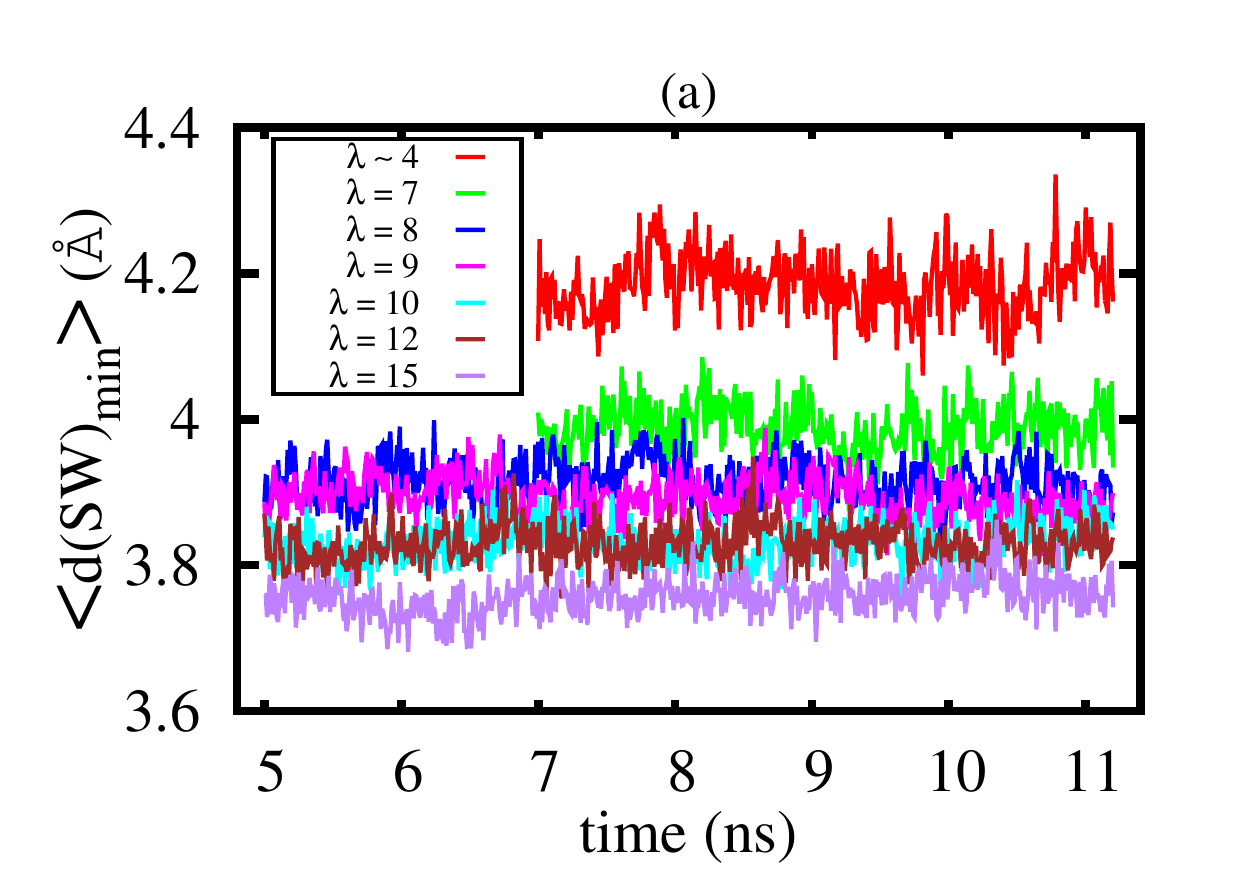}
  \includegraphics[scale= .64]{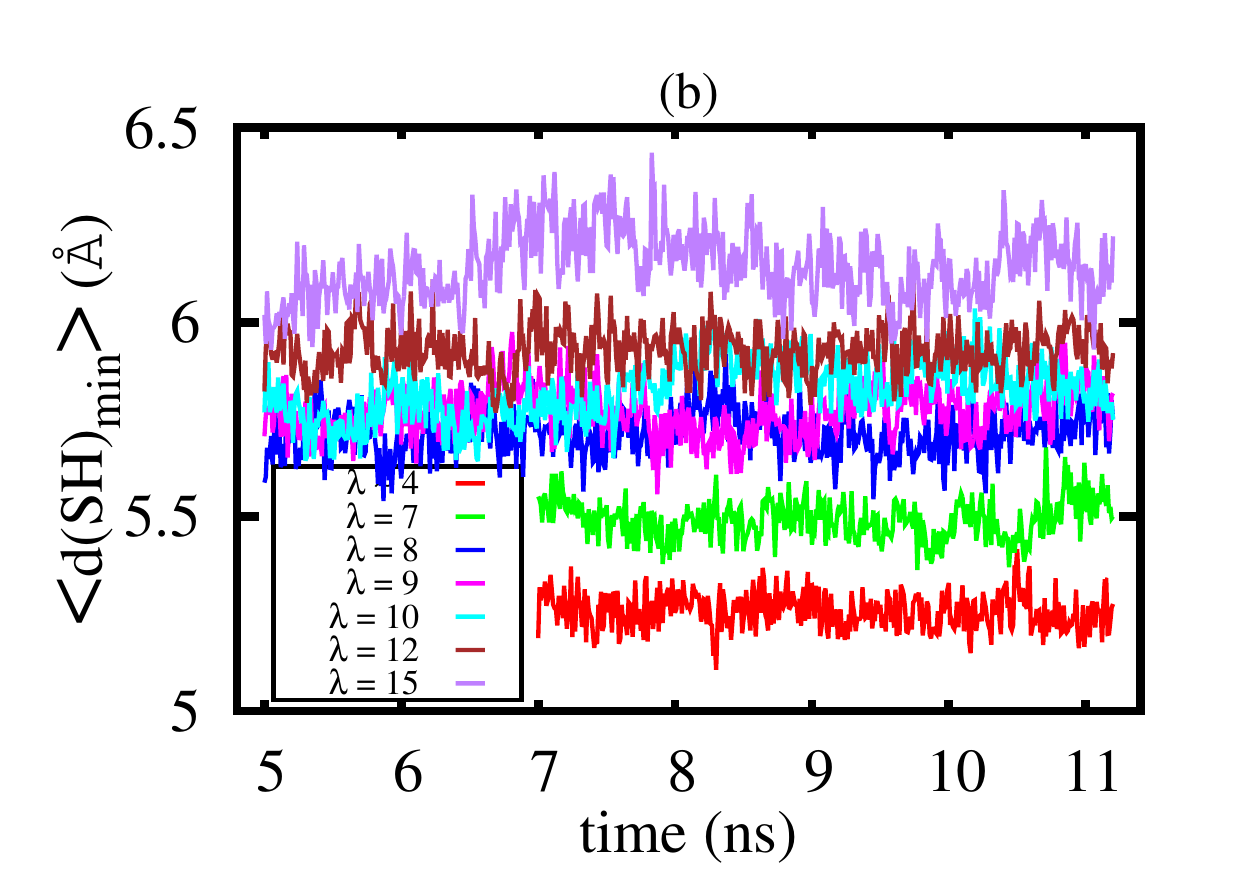}
  \caption{Average distance between a sulfur atom and its closest (a) water molecule and (b) hydronium ion, as a function of simulation time, for all hydration levels under study at 400 K.} 
   \label{fgr:min_d}
\end{figure}

As shown in Figure~\ref{fgr:min_d}(a), $\langle${d(SW)$_{\text{min}}$}$\rangle$ was found to decrease with increasing hydration: from around 4.2 \AA \ at $\lambda$ $\sim$ 4 to around 3.7 \AA \ at $\lambda$ = 15, decreasing slowly beyond $\lambda$ = 8. These distances are in the rage of the first peak of the SW RDF (Figure~\ref{fgr:swh_rdf}(a)), indicating various possible distances of the hydrogen bonded water molecules, within the first hydration shell of the sulfonate groups. The fluctuations in $\langle${d(SW)$_{\text{min}}$}$\rangle$ also decreased noticeably with increasing $\lambda$, indicating the water molecules to get tightly bound to the sulfonate groups with increasing water content.

In experimental studies, based on dielectric spectroscopy, it was predicted that water molecules should be more tightly bound to the sulfonate groups in sPEEK as compared to that in Nafion \citep{Paddison_MWDS}. Our observation that $\langle${d(SW)$_{\text{min}}$}$\rangle$ decreases with increasing $\lambda$ for sPEEK is in conformity with the above experimental observation of strong binding, especially when contrasted to increase in $\langle${d(SW)$_{\text{min}}$}$\rangle$ with hydration reported for Nafion \citep{Devnathan_nafion_1}.

Unlike $\langle${d(SW)$_{\text{min}}$}$\rangle$, $\langle${d(SH)$_{\text{min}}$}$\rangle$ (Figure~\ref{fgr:min_d}(b)) was found to increase with increasing $\lambda$ and so also the corresponding fluctuation. For all $\lambda$, $\langle${d(SH)$_{\text{min}}$}$\rangle$ was found to be larger than the position of first minima of SH RDF.  These observations suggest a weaker interaction between hydronium-sulfonate than that between water-sulfonate in sPEEK membranes. The SH RDF and $\langle${d(SH)$_{\text{min}}$}$\rangle$ indicate a large proportion of hydronium ions to occupy the second shell of sulfonate group, with a much lesser proportion in the first shell. Increase in $\lambda$ leads to solvation of the sulfonate group, thereby depleting hydronium ions from the first shell. Those in the second shell remain unaffected, and thus must already be a part of the water domains in the membrane. With increasing hydronium ion-sulfonate group distance, the binding becomes weaker leading to increased fluctuation in $\langle${d(SH)$_{\text{min}}$}$\rangle$. This behavior of SH RDF and $\langle${d(SH)$_{\text{min}}$}$\rangle$ is in contrast to that observed in Nafion  \citep{Devnathan_nafion_1}, where the hydronium ions possess stronger interaction with the sulfonate groups.
\subsection{Effect of hydration on transport properties}

\begin{figure}
  \includegraphics[scale= .64]{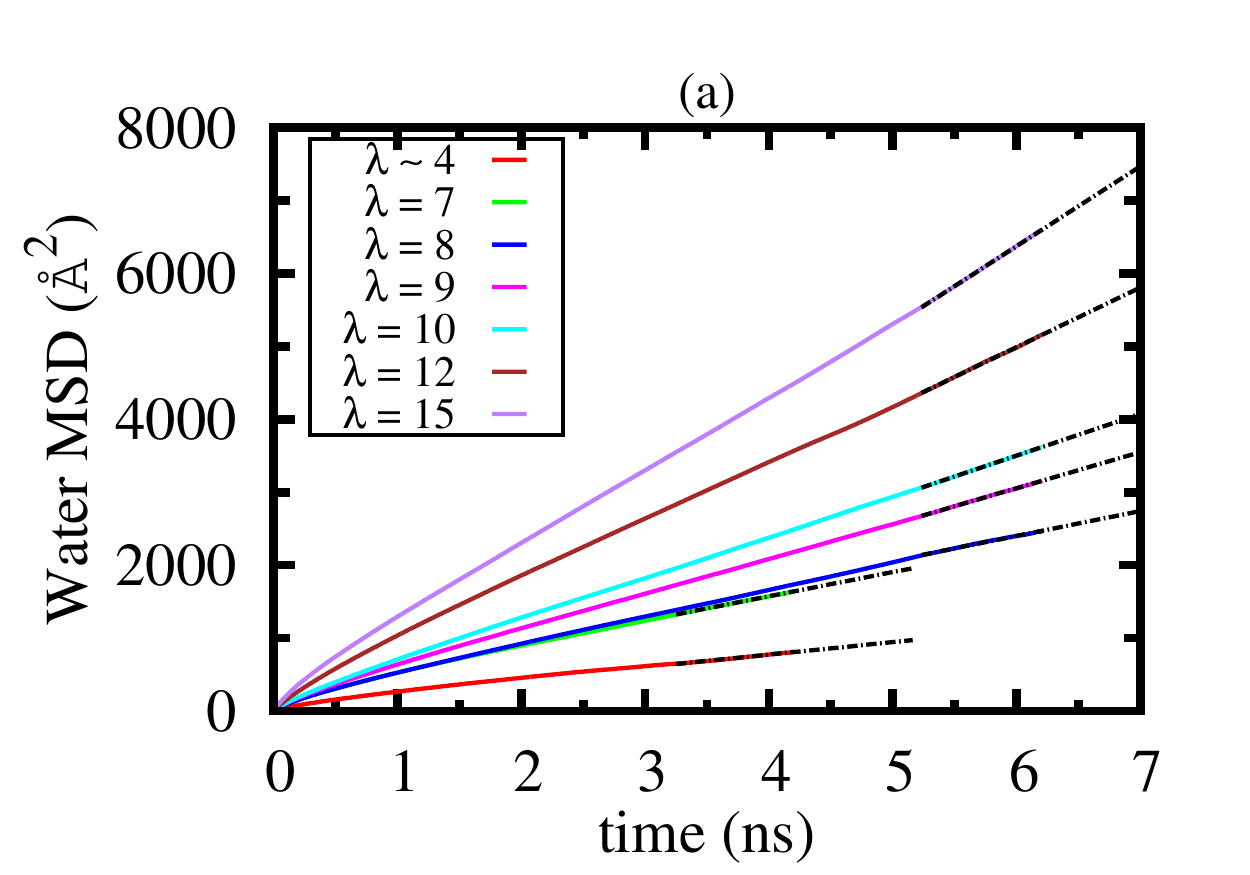}
  \includegraphics[scale= .64]{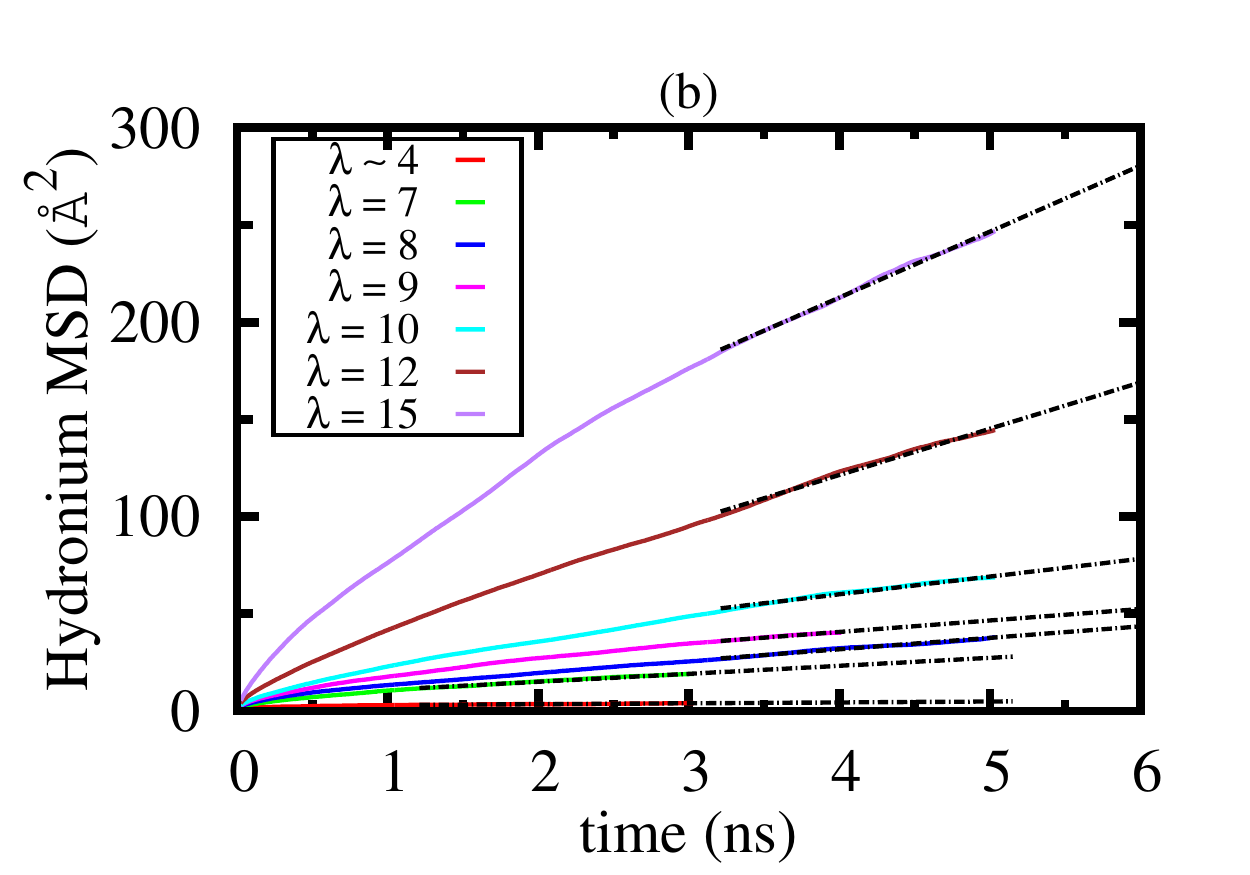}
  \caption{Mean-squared displacement of (a) water molecules and (b) hydronium ions in the sPEEK membrane at various hydration levels ($\lambda$) at 400 K. Dashed lines correspond to the long time limit slopes used for calculating diffusion coefficients.} 
   \label{fgr:msd_w_h}
\end{figure}

\begin{figure}
  \includegraphics[scale= .62]{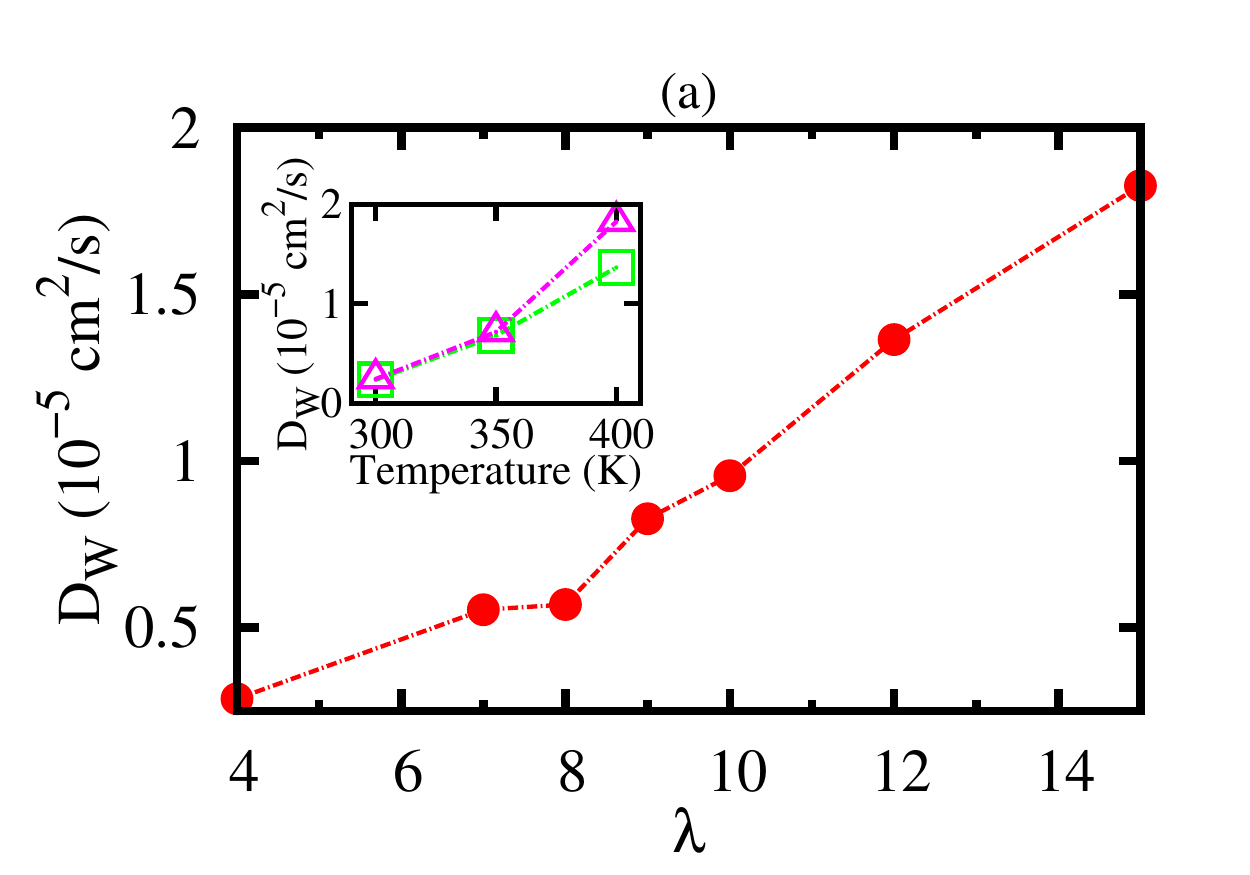}
  \includegraphics[scale= .62]{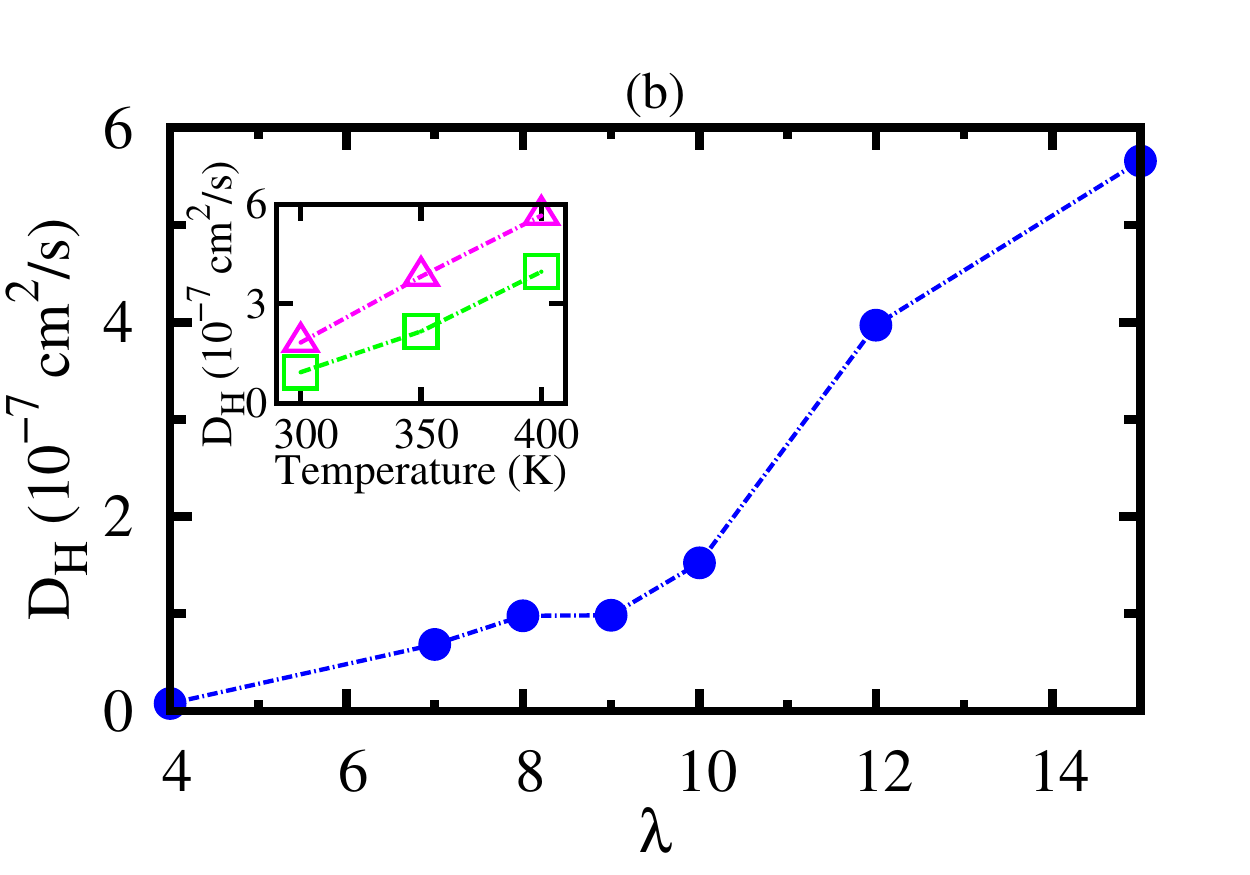}
  \caption{Change in the diffusion  coefficients of (a) water molecule (filled red circles) and (b) hydronium ion (filed blue circles) with varying $\lambda$ at 400 K. Insets show the variation of the diffusion  coefficients with temperature for $\lambda$ = 12 (open green squares) and $\lambda$ = 15 (open magenta triangles). Lines are guide to the eyes.} 
   \label{fgr:diffusion_coefficient}
\end{figure}

To investigate the effect of hydration on transport properties of water molecules and hydronium ions in sPEEK membrane, we calculated their MSDs. The slope of MSD for water (Figure~\ref{fgr:msd_w_h}(a)) was observed to steadily increase with $\lambda$, due to the development in the internal organization of the water phase in the membrane, as observed earlier for sPEEK by Brunello et al. \citep{Brunello_1} and for Nafion by Devanathan et al. \citep{Devnathan_nafion_2}. The diffusion coefficients were calculated from the corresponding MSDs as $D = \lim_{t\to\infty} \frac{MSD}{6t}$. The dashed lines in Figure~\ref{fgr:msd_w_h} show the range of data used for fitting the long time linear part of the MSDs to calculate the diffusion coefficients. The variations in the diffusion coefficients for water molecule (D$_\text{W}$) and that for hydronium ion (D$_\text{H}$) are plotted in Figure~\ref{fgr:diffusion_coefficient}. D$_\text{W}$ was found to increase with hydration from a value of $0.29 \times 10^{-5}$  cm$^2$ s$^{-1}$ at $\lambda$ $\sim$ 4 to $1.83  \times 10^{-5}$  cm$^2$ s$^{-1}$ at $\lambda$ = 15. D$_\text{W}$ at $\lambda$ = 15 (the maximum hydration level under study)  was found to be less than one-sixth of  the experimental diffusivity of bulk water at 400 K \citep{water_D}. This observation is in line with the experimental findings of Paddison et al. \citep{Paddison_MWDS} who suggested that large water uptake does not lead to bulk water phase in sPEEK membranes, as opposed to Nafion. The gradual increase in D$_\text{W}$ is similar to that observed in case of phenylated sPEEKK \citep{Devnathan_phspeekk}, and the corresponding values are comparable \citep{Mahajan_2, Brunello_1,  Devnathan_phspeekk}, though somewhat lower for sPEEK owing to the better water channel network in the former as compared to sPEEK.

The MSD of  hydronium, as shown in Figure~\ref{fgr:msd_w_h}(b),  indicate faster increase in hydronium mobility  at intermediate $\lambda$ values, as opposed to the gradual increase in case of water. Such a rapid increase in hydronium mobility has also been observed for Nafion \citep{Devnathan_nafion_2} earlier.  D$_\text{H}$ was found to increase from about $0.76 \times 10^{-8}$  cm$^2$ s$^{-1}$ at $\lambda$ $\sim$ 4 to $5.66  \times 10^{-7}$  cm$^2$ s$^{-1}$ at $\lambda$ = 15. The corresponding values were found to be comparable to those reported for sPEEK by Mahajan and Ganesan  \citep{Mahajan_2}, while lesser than those for phenylated sPEEKK \citep{Devnathan_phspeekk}, as expected. A pronounced increase in the value for D$_\text{H}$ was observed at $\lambda$ $\ge$ 10. The sudden increase in the  mobility of hydronium ions can be attributed to couple of reasons (a) their detachment from the sulfonate groups  and (b) formation of continuous water network in the membrane at higher water content, which provides a well connected path for the hydronium ions to diffuse through. We will come back to this point again later. For Nafion membrane, a similar jump in D$_\text{H}$  was observed in experiment at $\lambda$ = 5 \citep{Nafion_expt_D} and in simulation between $\lambda$ = 5 and 7 \citep{Devnathan_nafion_2, Devnathan_nafion_3}.

At sufficiently low hydration, hydronium ions remain bound to the sulfonate groups and therefore, their mobility is hindered. As we have seen earlier in our analysis in Section 3.1, as $\lambda$ increases, the hydronium ions get unbound from the sulfonate groups and move into the growing water phase. At the onset of percolation, when a connected water network gets formed, the mobility of hydronium ions suddenly increases. Therefore, it can be concluded that the sharp increase in D$_\text{H}$ is due to percolation.

The insets of Figure~\ref{fgr:diffusion_coefficient} show the variation of D$_\text{W}$ and D$_\text{H}$ with temperature for hydration levels 12 and 15. While both D$_\text{W}$ and D$_\text{H}$ increased monotonously with temperature, the increase in $\lambda$ did not lead to as  appreciable a change in the temperature dependence of D$_\text{W}$ as in D$_\text{H}$, at low temperatures. Similar observations were reported earlier for phenylated sPEEKK \citep{Devnathan_phspeekk}, with no significant change in the nanophase segregated structure or swelling in the membrane between 300 K and 400 K, but enhanced dynamics of water molecules and hydronium ions. 


\subsection{Cluster analysis}

The collective water phase in sPEEK membrane and the effect of hydration on the same was characterized  in terms of  clustering of water molecules and  hydronium ions. Clusters were identified using a distance based neighborhood criteria with a cutoff of 3.5 \AA.   The cutoff distance was chosen based on our earlier  reference bulk  liquid water simulation \citep{cg_speek_madhu} using the modified TIP3P water model \citep{TIP3P}. The same cutoff distance criteria has also been used to analyze water clusters in  earlier works on Nafion \citep{Devnathan_nafion_3} and phenylated sPEEKK \citep{Devnathan_phspeekk}. We study the variation in number and size of the clusters and their distributions as the water content in the membrane increases. In the sections below, we discuss these results in brief.

\subsubsection{Variation in number and size of clusters with hydration}

\begin{figure}
\centering
  \includegraphics[scale= .66]{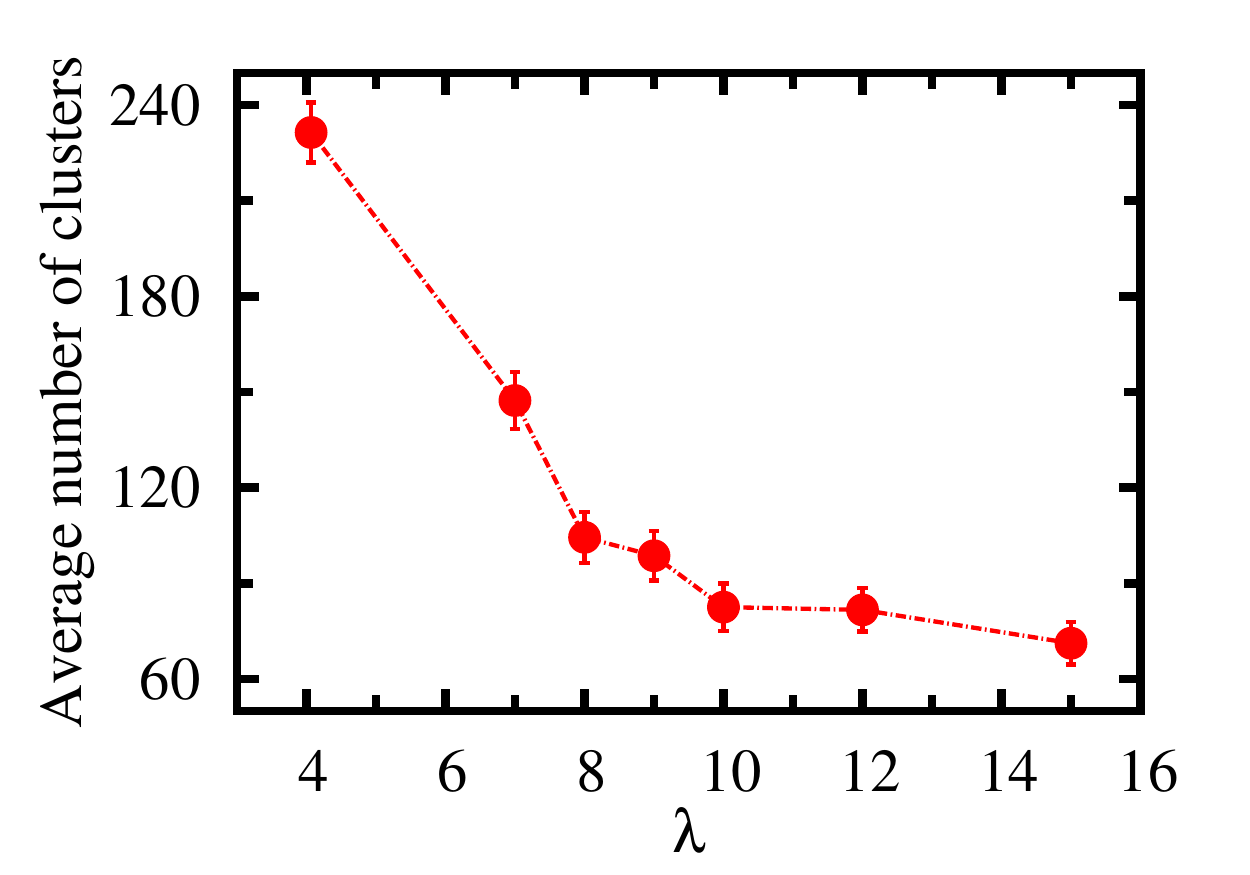}
  \caption{Average number of clusters as a function of  hydration level  ($\lambda$)  at 400 K. Error bars indicate the corresponding standard deviations. Line is guide to the eyes.} 
   \label{fgr:num_cluster}
\end{figure}

In Figure~\ref{fgr:num_cluster}, we plot the total number of clusters, averaged over the production trajectory for various hydration levels, along with the corresponding standard deviations. There was a sharp decrease in the number up to   $\lambda$ = 10, beyond which, the change was observed to be gradual.   At low water content, sPEEK is known to possess  small dispersed water clusters \citep{Mahajan_1, Brunello_1, Komarov_1}. As $\lambda$  increases, these clusters grow by merging, leading to decrease in the average number of clusters. The fluctuation in the number of clusters was also found to decrease with increasing $\lambda$, suggesting the transient clusters to become  stable as $\lambda$ increases.

To further investigate the water phase, we calculated the size of the largest cluster  (S$_\text{max}$) for various hydration levels. As summarized in Table~\ref{tbl:max_cluster_size}, (S$_\text{max}$) was found to increase with increasing $\lambda$. While less than 15\% of available water and hydronium constituted the largest cluster at $\lambda$ $\sim$ 4,  more than 95\% of them got clustered at $\lambda$ = 10. Beyond this point, S$_\text{max}$ increased gradually, with isolated clusters still present at $\lambda$ = 15. Such isolated clusters can lead to the occurrence of dead ends in water channels in sPEEK membranes, which have been observed in scattering experiments \cite{Kreuer}. In  atomistic simulation on Nafion \citep{Devnathan_nafion_3} by Devanathan et al., a spanning water network was observed beyond $\lambda$ = 5, where all the water molecules and hydronium ions belonged to a single cluster. Also, the average number of clusters decreased nearly an order of magnitude between $\lambda$ = 3 and 7.  Our results therefore, are in agreement with earlier experimental observations based on scattering studies \citep{Kreuer}, which suggested the extent of phase separation in sPEEK to be lesser than that in Nafion. While the atomistic simulations of Nafion referred here contained only four polymer chains with total 40 sulfonate groups, our sPEEK system containing 240 sulfonate groups is comparatively much larger. 

In the results on Nafion, the eventual number of clusters beyond percolation was shown to be one. On the other hand, we show that the number of clusters in sPEEK remain  of the order of few tens to nearly a hundred, indicating the presence of isolated clusters even after percolation. An important aspect of our claim, that this observation and other sPEEK results are in conformity with the experimental observations, is based on the fact that our simulation system is reasonably large, and therefore  more clearly represents the large length scales required for phase separation and clustering.

 \begin{widetext}
 \begin{center}
\begin{table}
\centering
  \caption{Size of the largest cluster for various hydration levels.}
  \label{tbl:max_cluster_size}
\begin{tabular}{c c c c}
\hline \hline
Hydration & Number of water molecules   & Maximum   & Percentage of \\ 
level  &  \& hydronium ions    & cluster size  & N$_\text{W}$ + N$_\text{H}$ in S$_\text{max}$  \\
 ($\lambda$) & (N$_\text{W}$ + N$_\text{H}$)   & (S$_\text{max}$) & (\%) \\
\hline
4.067 & 1216 & 179 & 14.72 \\
7 & 1920 & 1587 & 82.66 \\
8 & 2160 & 1957 & 90.60 \\
9 & 2400 & 2240 & 93.33 \\
10 & 2640 & 2520 & 95.45 \\
12 & 3120 & 3004 & 96.28 \\
15 & 3840 & 3753 & 97.73 \\
\hline
\hline
\end{tabular} 
\end{table}

\begin{figure}
\centering
  \includegraphics[scale= .3]{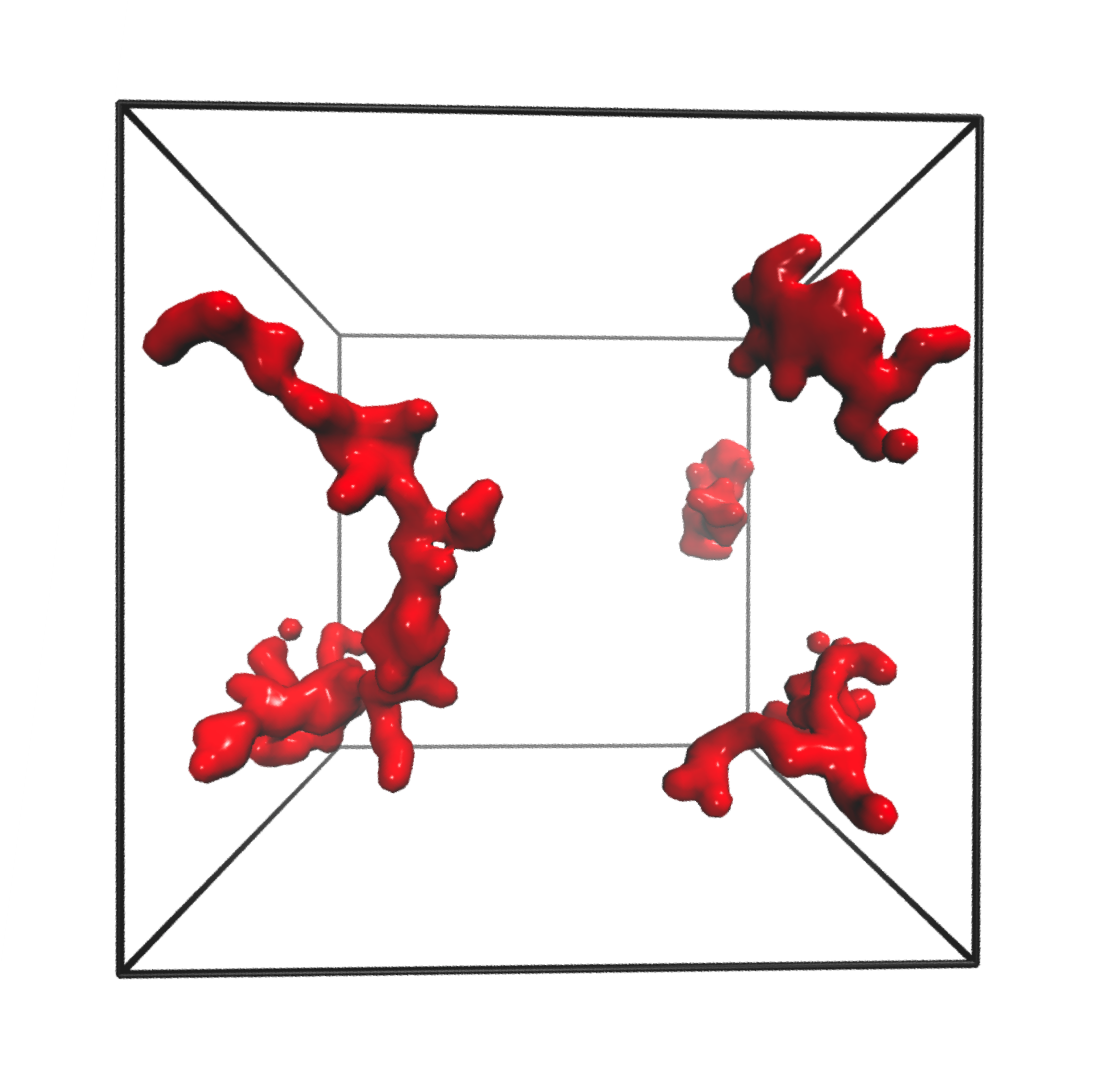}
  \includegraphics[scale= .3]{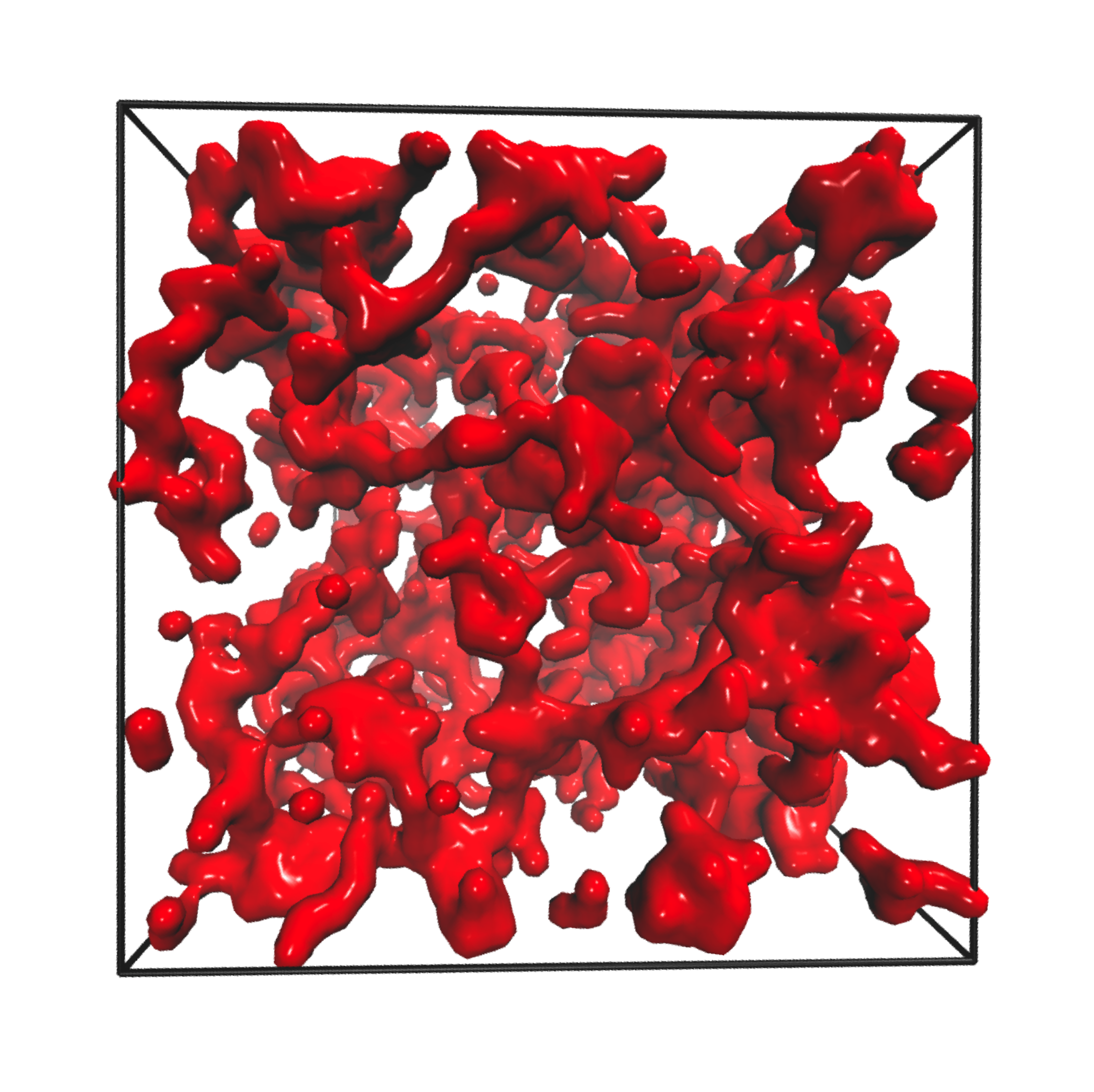}
  \includegraphics[scale= .3]{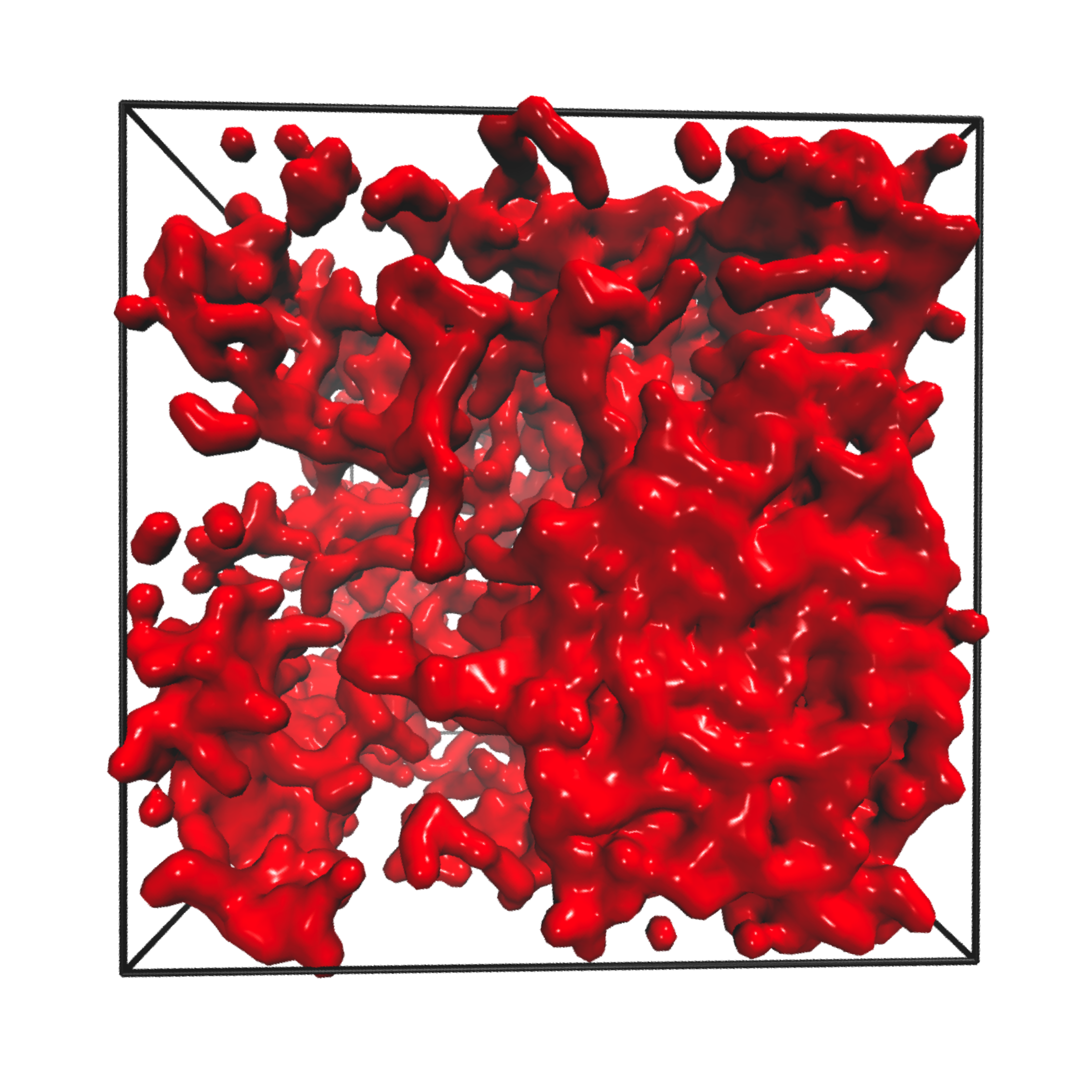}  
  \caption{Water molecules and hydronium ions in the simulation box belonging to the largest cluster for hydration levels ($\lambda$): 4.067, 10, and 15 (from left to right). Only the oxygen atoms are shown for clarity. The snapshots have been rendered using the visualization tool, VMD \citep{VMD} and the iso-surface has been generated using QuickSurf \citep{quicksurf}. }
   \label{fgr:max_cluster_snapshot}
\end{figure}
 \end{center}

\end{widetext}

Snapshots of the simulation boxes containing the largest clusters for hydration levels 4.067, 10, and 15 (as given in Table~\ref{tbl:max_cluster_size}) are shown in Figure~\ref{fgr:max_cluster_snapshot}. The largest cluster at $\lambda$ $\sim$ 4 is found to occupy a small fraction of the simulation box volume and forms a narrow connected path of water molecules and hydronium ions. A well connected network, spanning the entire simulation box, is observed at $\lambda$ = 10, with the appearance of larger water domains. These domains further grow in size,  leading to the formation of the nano-meter-size water domains (or channels), that have been observed in scattering experiments \citep{Kreuer,  Mauritz, Schmidt-Rohr}.  Such a large water domain is observed to  span more than half of the simulation box length ($>$ 3 nm) at $\lambda$ = 15.
\subsubsection{Cluster size distribution and mean cluster size}

\begin{figure}
  \includegraphics[scale= .68]{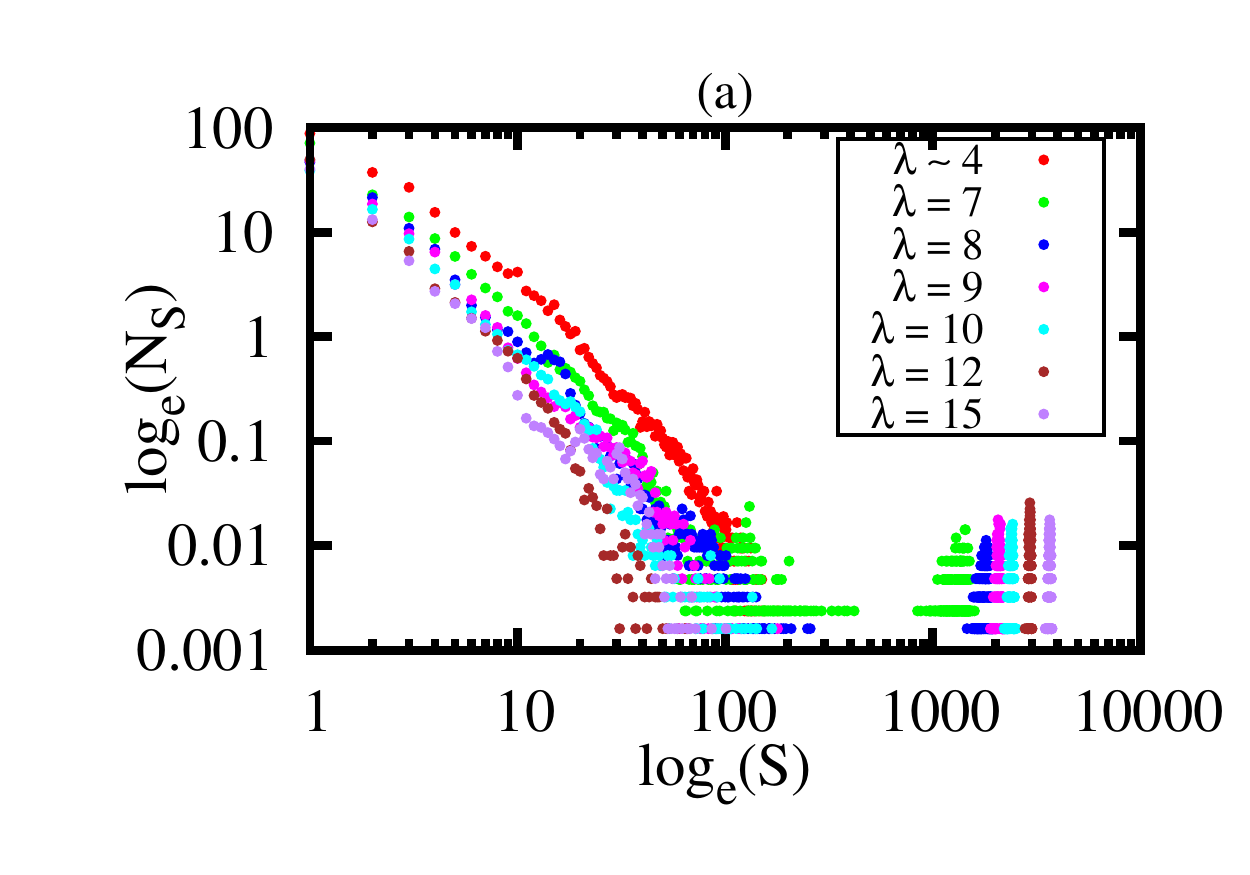}
  \includegraphics[scale= .68]{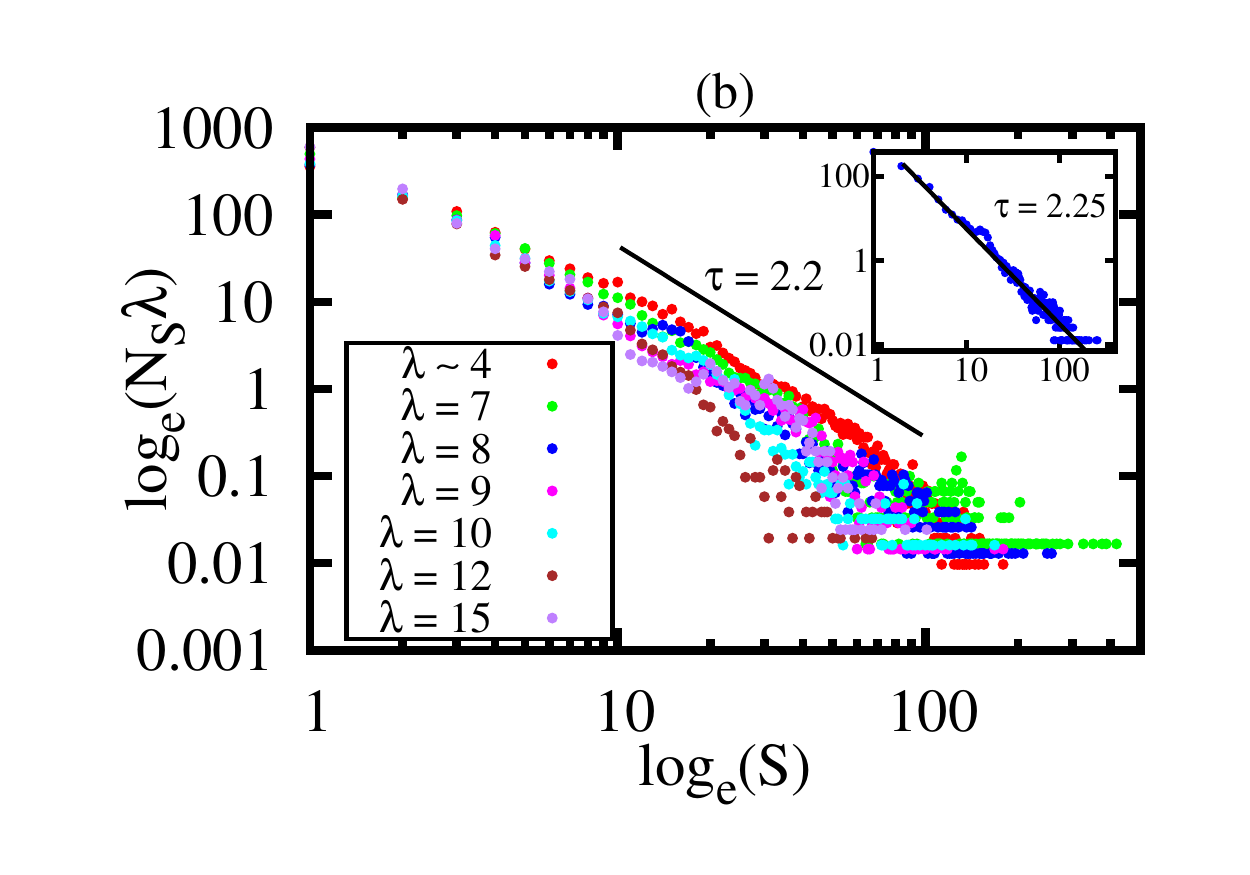}
  \caption{(a) Distribution of cluster sizes for various hydration levels under study at 400 K. (b) Distributions over the region of smaller cluster sizes. The line represents a power-law dependency with slope $\tau$ = 2.2 for three-dimensional percolation. The inset in (b) shows the distribution of cluster sizes for $\lambda$ = 8 and the line represents a power-law dependency with slope $\tau$ = 2.25.} 
   \label{fgr:cluster_distribution}
\end{figure}

To get further insight into the water phase in sPEEK membrane, we calculated the cluster size distribution for all the hydration levels under study. Figure~\ref{fgr:cluster_distribution}(a), shows the variation in N$_\text{S}$, the occurrence probability (number of clusters with size S), with hydration.  Isolated water molecules or hydronium ions were considered as clusters of size S = 1. The distribution was found to  span over two orders of magnitude in S at $\lambda$ $\sim$ 4 to more than three orders at $\lambda$ = 15.  The two separate regimes in N$_\text{S}$, comprising of continuously varying cluster sizes at lower S and isolated large clusters at higher S, became more distinct as  $\lambda$ increased beyond 7, suggesting the onset of spanning clusters at $\lambda$ = 7. In three dimensions, at the onset of random percolation, N$_\text{S}$ is known to obey a power law \citep{random_percolation}: N$_\text{S}$ $\sim$ S$^{-2.2}$. This behavior  over the lower range of S is shown  in Figure~\ref{fgr:cluster_distribution}(b), where the expected power law  is shown with a line as   guide to the eye.  The  occurrence probability, N$_\text{S}$ is multiplied by the corresponding $\lambda$ for better comparison.

It should be noted that the threshold to percolation is known to be system size dependent \citep{Stauffer}. Additionally, the uncertainty  in the  onset of percolation is larger when the system size is smaller  \citep{percolation_protein}. In our simulations, the largest extent of power law was obtained at   $\lambda$ = 8, with the entire small S range represented by a power law exponent of 2.25, as shown in the inset of  Figure~\ref{fgr:cluster_distribution}(b). Based on the agreement with the power law and the similarity of the exponent with that for random percolation, we consider $\lambda$ = 8 to be closest to percolation threshold in sPEEK membranes for the system sizes considered here.   In a similar analysis for Nafion \citep{Devnathan_nafion_3}, the authors reported $\lambda$ =  5 to be closest to percolation threshold as the widest range of distribution data (S = 2 to 199) could be fit to a power law with an exponent of 1.6.

\begin{figure}
\centering
  \includegraphics[scale= .66]{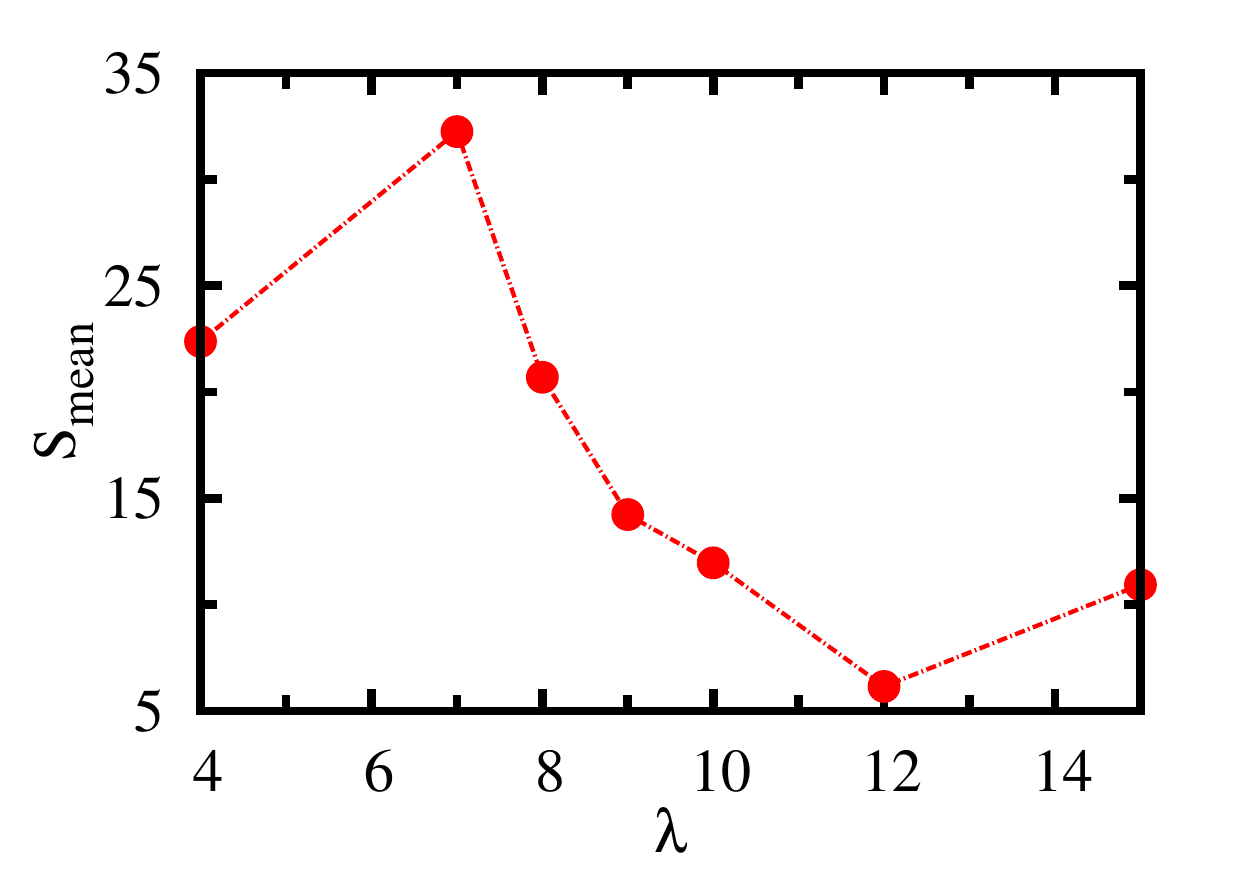}
  \caption{Mean cluster size S$_\text{mean}$ as a function of $\lambda$ at 400 K. Line is guide to the eyes.} 
   \label{fgr:mean_cluster_size}
\end{figure}

We also  calculated the mean cluster size S$_\text{mean}(\lambda)= \frac{\sum \text{N}_\text{S} \text{S}^2}{\sum \text{N}_\text{S} \text{S}},$ where the sum is over all clusters excluding second mode in the distribution (plotted in Figure~\ref{fgr:mean_cluster_size}).  S$_\text{mean}(\lambda)$ is known to exhibit  a maximum just below percolation  in finite size systems \citep{Stauffer, percolation_dna}. The maximum at $\lambda$ = 7 suggests percolation to occur after this hydration, which is in agreement with the analysis of the cluster size distribution.

Figure~\ref{fgr:system_snapshot} shows the atomistic snapshots of the sPEEK membrane at $\lambda$: 4.067, 10, and 15. Small isolated water clusters present at $\lambda$ $\sim$ 4 can be seen to grow and get eventually connected at $\lambda$ = 10. Beyond this hydration, the water cluster grow to form larger domains. Large water domains connected by narrower networks of water are clearly visible at $\lambda$ = 15. The sulfur atoms scattered in the polymer matrix at low $\lambda$, lie at the boundary of the water phase at higher $\lambda$, while the hydronium ions are observed to be in the continuous water phase. In line with our analysis, $\lambda$ = 10 seems to be beyond percolation with a well connected water phase. $\lambda$ = 15 clearly has well-developed water network and well-defined water domains. However, the presence of isolated clusters can be observed even at $\lambda$ = 15.

Our simulation results indicate random percolation transition,  in the sPEEK membrane under study, to occur between $\lambda$ = 8 and 10. This is in agreement with earlier investigation on sPEEK using density functional theory by Komarov et al. \citep{Komarov_2}, who  reported a percolation threshold hydration level of $\sim$ 9. A value of 7.92 was reported as percolation threshold in phenylated sPEEKK by Devanathan et al. \citep{Devnathan_phspeekk}.

\begin{widetext}
 \begin{center}
\begin{figure}
\centering
  \includegraphics[scale= .25]{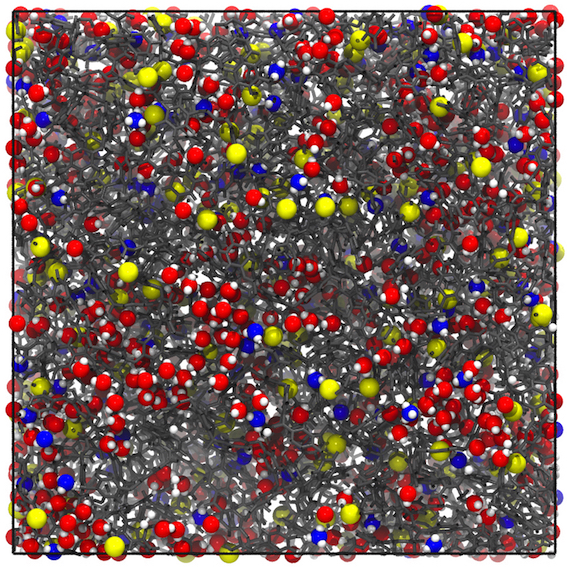}
  \includegraphics[scale= .25]{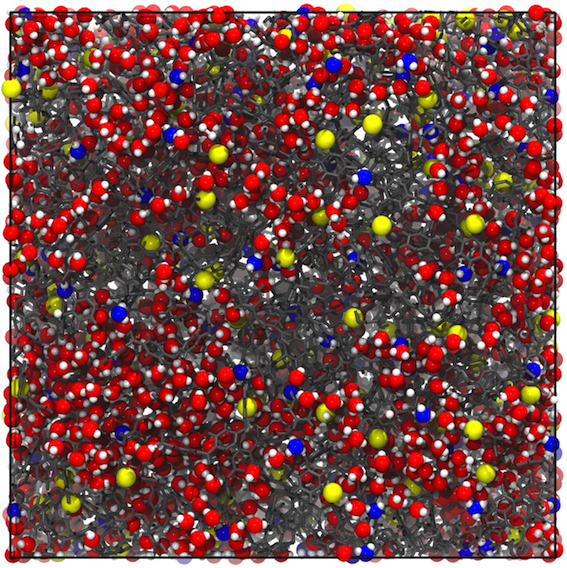}
  \includegraphics[scale= .256]{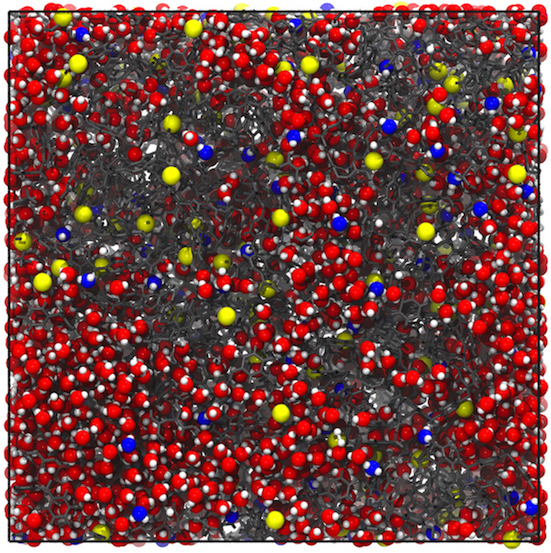}  
  \caption{Atomistic snapshots of sPEEK membrane for hydration levels  ($\lambda$): 4.067, 10,  and  15 (from left to right). Sulfur atoms, water oxygen, hydronium oxygen, and hydrogen atoms are shown in yellow, red, blue, and white respectively as van der Waals spheres. Polymer backbones are shown in grey using bond representation. The snapshots have been rendered using VMD \citep{VMD}. }
   \label{fgr:system_snapshot}
\end{figure}
 \end{center}
\end{widetext}

\section{SUMMARY AND CONCLUSION}

This work has focused on the effect of hydration on the structural and dynamical properties of sPEEK membranes. We systematically investigate the effect of increasing water content: from the  level of individual molecules, in terms of their local structuring and transport, to collective properties, in terms of the growing water phase in the membrane. A combination of coarse-graining and back-mapping methodology is used to  access the relevant length scales in atomistic simulation.

At low hydration, small isolated water clusters are present in the membrane and hydronium ions stay bound to the nearby sulfonate groups. As hydration increases, water molecules solvate the sulfonate group, thereby detaching the hydronium ions from the sulfonate groups and pushing them into the growing water phase. Water molecules are found to be more strongly bound to the sulfonate groups in sPEEK than in Nafion, as previously suggested by experiments \citep{Paddison_MWDS}.  With increasing hydration, the isolated clusters merge to form larger connected network. However, unlike Nafion, where all available  water molecules and hydronium ions in the system form a spanning cluster at $\lambda$ as low as 5 \citep{Devnathan_nafion_3}, small clusters can remain isolated in sPEEK even at $\lambda$ = 15. This is in agreement with earlier experimental observations suggesting the extent of phase separation in sPEEK to be lesser than that in Nafion \citep{Kreuer}. The mobility of water molecules remains much less than that of bulk water, even at the highest water content, demonstrating the bound nature of water in sPEEK membranes, in agreement with earlier experimental findings \citep{Paddison_MWDS}.

Further analyses suggest a percolation transition occurring in the sPEEK membrane between $\lambda$ = 8 and 10, where the mobility of hydronium ion suddenly increases.  This sudden jump in the diffusion coefficient  can be attributed to the well connected water network forming beyond percolation, which promotes the hydronium diffusion through the membrane.

On the methodology front, we show that equilibrated structure from multi-chain CG simulation, with well-mixed near-equilibrium polymer configuration, can be used as a starting point for further atomistic simulations on hydrated sPEEK. This method leads to faster equilibration of the atomistic system, without compromising the equilibrium structural properties. Environmental condition such as hydration and temperature can be varied at this point to study their effect on the static and dynamic properties of the system. Such a protocol can accelerate the equilibration of large-scale atomistic polymer system without the need of sophisticated computational technique or resource.

\section*{Acknowledgements} 

The computation for this work has been carried  out at High Performance Computing Environment at IIT Madras.



\begin{thebibliography}{111}

\bibitem{Mauritz}Mauritz, K. A.; Moore, R. B. State of Understanding of Nafion. {\it Chem. Rev.} {\bf 2004}, 104, 4535–4586.

\bibitem{speek_thermal_stability}Swier, S.; Chun, Y. S.; Gasa, J.; Shaw, M. T.; Weiss, R. Sulfonated poly(ether ketone ketone) ionomers as proton exchange membranes. {\it Polym Eng. Sci.} {\bf 2005}, 45, 1081–1091.

\bibitem{high_temp_speek}Lakshmanan, B.; Huang, W.; Olmeijer, D.; Weidner, J. W. Polyetheretherketone Membranes for Elevated Temperature PEMFCs. Electrochem. {\it Solid State Lett.} {\bf 2003}, 6,A282–A285.

\bibitem{Kreuer}Kreuer, K. D. On the Development of Proton Conducting Polymer Membranes for Hydrogen and Methanol Fuel Cells. {\it J. Membr. Sci.} {\bf 2001}, 185, 29–39.

\bibitem{Schmidt-Rohr}Schmidt-Rohr, K.; Chen, Q. Parallel cylindrical water nanochannels in Nafion fuel-cell membranes. {\it Nat. Mater.} {\bf 2008}, 7, 75–83.

\bibitem{DPD_amphiphilic}Dorenbos, G.; Morohoshi, K. Percolation thresholds in hydrated amphiphilic polymer membranes. {\it J. Mater. Chem.} {\bf 2011}, 21, 13503–13515.

\bibitem{DPD_nafion}Vishnyakov, A.; Neimark, A. V. Self-Assembly in Nafion Membranes upon Hydration: Water Mobility and Adsorption Isotherms. {\it J. Phys. Chem. B} {\bf 2014}, 118, 11353–11364.

\bibitem{Wescott_Nafion}Wescott, J. T.; Qi, Y.; Subramanian, L.; Weston Capehart, T. Mesoscale simulation of morphology in hydrated perfluorosulfonic acid membranes. {\it J. Chem. Phys.} {\bf 2006}, 124, 134702.

\bibitem{DPD_speek}Komarov, P. V.; Veselov, I. N.; Khalatur, P. G. Self-organization of amphiphilic block copolymers in the presence of water: A mesoscale simulation. {\it Chem. Phys. Lett.} {\bf 2014}, 605-606, 22 – 27.

\bibitem{Komarov_1}Komarov, P. V.; Veselov, I. N.; Chu, P. P.; Khalatur, P. G. Mesoscale Simulation of Polymer Electrolyte Membranes Based on Sulfonated Poly(Ether Ether Ketone) and Nafion. {\it Soft Matter} {\bf 2010}, 6, 3939–3956.

\bibitem{Komarov_2}Komarov, P. V.; Veselov, I. N.; Chu, P. P.; Khalatur, P. G.; Khokhlov, A. R. Atomistic and Mesoscale Simulation of Polymer Electrolyte Membranes Based on Sulfonated Poly(Ether Ether Ketone). {\it Chem. Phys. Lett.} {\bf 2010}, 487, 291–296.

\bibitem{Mahajan_1}Mahajan, C. V.; Ganesan, V. Atomistic Simulations of Structure of Solvated Sulfonated Poly(Ether Ether Ketone) Membranes and Their Comparisons to Nafion: I. Nanophase Segregation and Hydrophilic Domains. {\it J. Phys. Chem. B} {\bf 2010}, 114, 8357–8366.

\bibitem{Mahajan_2}Mahajan, C. V.; Ganesan, V. Atomistic Simulations of Structure of Solvated Sulfonated Poly(ether ether ketone) Membranes and Their Comparisons to Nafion: II. Structure and Transport Properties of Water, Hydronium Ions, and Methanol. {\it J. Phys. Chem. B} {\bf 2010}, 114, 8367–8373.

\bibitem{Devnathan_nafion_1}Devanathan, R.; Venkatnathan, A.; Dupuis, M. Atomistic Simulation of Nafion Membrane: I. Effect of Hydration on Membrane Nanostructure. {\it J. Phys. Chem. B} {\bf 2007}, 111, 8069–8079.

\bibitem{Devnathan_nafion_2}Devanathan, R.; Venkatnathan, A.; Dupuis, M. Atomistic Simulation of Nafion Membrane. 2. Dynamics of Water Molecules and Hydronium Ions. {\it J. Phys. Chem. B} {\bf 2007}, 111, 13006–13013.

\bibitem{Devnathan_nafion_3}Devanathan, R.; Venkatnathan, A.; Rousseau, R.; Dupuis, M.; Frigato, T.; Gu, W.; Helms, V. Atomistic Simulation of Water Percolation and Proton Hopping in Nafion Fuel Cell Membrane. {\it J. Phys. Chem. B} {\bf 2010}, 114, 13681–13690.

\bibitem{Devnathan_phspeekk}Devanathan, R.; Idupulapati, N.; Dupuis, M. Molecular modeling of the morphology and transport properties of two direct methanol fuel cell membranes: Phenylated sulfonated poly(ether ether ketone ketone) versus Nafion. {\it J. Mater. Res.} {\bf 2012}, 27, 1927–1938.

\bibitem{Brunello_1}Brunello, G.; Lee, S. G.; Jang, S. S.; Qi, Y. A Molecular Dynamics Simulation Study of Hydrated Sulfonated Poly(Ether Ether Ketone) for Application to Polymer Electrolyte Membrane Fuel Cells: Effect of Water Content. {\it J. Renew. Sustain. Energy} {\bf 2009}, 1, 033101.

\bibitem{Brunello_2}Brunello, G. F.; Mateker, W. R.; Lee, S. G.; Choi, J. I.; Jang, S. S. Effect of temperature on structure and water transport of hydrated sulfonated poly(ether ether ketone): A molecular dynamics simulation approach. {\it J. Renew. Sustain. Energy} {\bf 2011}, 3 .

\bibitem{Bahlakeh2012}Bahlakeh, G.; Nikazar, M.; Hafezi, M.-J.; Dashtimoghadam, E.; Hasani-Sadrabadi, M. M. Molecular dynamics simulation study of proton diffusion in polymer electrolyte membranes based on sulfonated poly (ether ether ketone). {\it Int. J. Hydrogen Energy} {\bf 2012}, 37, 10256 – 10264.

\bibitem{Karo_large_system}Karo, J.; Aabloo, A.; Thomas, J. O.; Brandell, D. Molecular Dynamics Modeling of Proton Transport in Nafion and Hyflon Nanostructures. {\it J. Phys. Chem. B} {\bf 2010}, 114, 6056–6064.

\bibitem{Karo_small_system}Brandell, D.; Karo, J.; Liivat, A.; Thomas, J. O. Molecular dynamics studies of the Nafion R , Dow R and Aciplex R fuel-cell polymer membrane systems. {\it J. Mol. Model.} {\bf 2007}, 13, 1039–1046.

\bibitem{Voth_large_aa}Knox, C. K.; Voth, G. A. Probing Selected Morphological Models of Hydrated Nafion Using Large-Scale Molecular Dynamics Simulations. {\it J. Phys. Chem. B} {\bf 2010}, 114, 3205–3218.

\bibitem{Komarov2013_large_aa}Komarov, P. V.; Khalatur, P. G.; Khokhlov, A. R. Large-scale atomistic and quantum- mechanical simulations of a Nafion membrane: Morphology, proton solvation and charge transport. {\it Beilstein J. Nanotechnol.} {\bf 2013}, 4, 567–587.

\bibitem{cg_speek_madhu}Tripathy, M.; Deshpande, A. P.; Kumar, P. B. S. How Much Can We Coarse-Grain while Retaining the Chemical Specificity? A Study of Sulfonated Poly(ether ether ketone). {\it Macromol. Theory Simul.} {\bf 2016}, 25, 155–169.

\bibitem{IBI}Reith, D.; P{\"{u}}tz, M.; M{\"{u}}ller-Plathe, F. Deriving Effective Mesoscale Potentials from Atomistic Simulations. {\it J. Comput. Chem.} {\bf 2003}, 24, 1624–1636.

\bibitem{Dreiding}Mayo, S. L.; Olafson, B. D.; Goddard, W. A. DREIDING: A Generic Force Field for Molecular Simulations. {\it J. Phys. Chem.} {\bf 1990}, 94, 8897–8909.

\bibitem{TIP3P}Price, D. J.; Brooks, C. L. A Modified TIP3P Water Potential for Simulation with Ewald Summation. {\it J. Chem. Phys.} {\bf 2004}, 121, 10096–10103.

\bibitem{Jang}Jang, S. S.; Molinero, V.; Cagin, T.; Goddard, W. A. Nanophase-Segregation and Transport in Nafion 117 from Molecular Dynamics Simulations: Effect of Monomeric Sequence. {\it J. Phys. Chem. B} {\bf 2004}, 108, 3149–3157.

\bibitem{Lammps}Plimpton, S. Fast Parallel Algorithms for Short-range Molecular Dynamics. {\it J. Comput. Phys.} {\bf 1995}, 117, 1–19.

\bibitem{high_t_pem}Zhang, J.; Xie, Z.; Zhang, J.; Tang, Y.; Song, C.; Navessin, T.; Shi, Z.; Song, D.; Wang, H.; Wilkinson, D. P. et al. High temperature PEM fuel cells. {\it J. Power Sources} {\bf 2006}, 160, 872 – 891.

\bibitem{swelling_saxs}G{\'{e}}bel, G. Structure of Membranes for Fuel Cells: SANS and SAXS Analyses of Sulfonated PEEK Membranes and Solutions. {\it Macromolecules} {\bf 2013}, 46, 6057–6066.

\bibitem{Huang}Huang, R. Y. M.; Shao, P.; Burns, C. M.; Feng, X. Sulfonation of Poly(Ether Ether Ketone)(PEEK): Kinetic Study and Characterization. {\it J. Appl. Polym. Sci.} {\bf 2001}, 82, 2651–2660.

\bibitem{Zaidi}Zaidi, S.; Mikhailenko, S.; Robertson, G.; Guiver, M.; Kaliaguine, S. Proton conducting composite membranes from polyether ether ketone and heteropolyacids for fuel cell applications. {\it J. Membr. Sci.} {\bf 2000}, 173, 17 – 34.

\bibitem{sulfate_anion}Cannon, W. R.; Pettitt, B. M.; McCammon, J. A. Sulfate Anion in Water: Model Structural, Thermodynamic, and Dynamic Properties. {\it J. Phys. Chem.} {\bf 1994}, 98, 6225–6230.

\bibitem{Paddison_MWDS}Paddison, S. J.; Bender, G.; Kreuer, K. D.; Nicoloso, N.; Zawodzinski, T. A. The microwave region of the dielectric spectrum of hydrated Nafion R and other sulfonated membranes. {\it J. New Mater. Electrochem. Syst.} {\bf 2000}, 3, 293–302.

\bibitem{VMD}Humphrey, W.; Dalke, A.; Schulten, K. VMD: Visual Molecular Dynamics. {\it J. Mol. Graphics} {\bf 1996}, 14, 33–38.

\bibitem{quicksurf}Krone, M.; Stone, J. E.; Ertl, T.; Schulten, K. Fast Visualization of Gaussian Density Surfaces for Molecular Dynamics and Particle System Trajectories. {\it EuroVis - Short Papers 2012} {\bf 2012}, 67–71.

\bibitem{random_percolation}Jan, N.; Stauffer, D. Random Site Percolation in Three Dimensions. {\it Int. J. Mod. Phys. C} {\bf 1998}, 09, 341–347.

\bibitem{Stauffer}Stauffer, D.; Aharony, A. Introduction to Percolation Theory; Taylor \& Francis, {\bf 1994}.

\bibitem{percolation_protein}Smolin, N.; Oleinikova, A.; Brovchenko, I.; Geiger, A.; Winter, R. Properties of Spanning Water Networks at Protein Surfaces. {\it J. Phys. Chem. B} {\bf 2005}, 109, 10995–11005.

\bibitem{percolation_dna}Brovchenko, I.; Krukau, A.; Oleinikova, A.; Mazur, A. K. Ion Dynamics and Water Percolation Effects in DNA Polymorphism. {\it J. Amer. Chem. Soc.} {\bf 2008}, 130, 121–131.

\bibitem{water_D}Lee, S. H. Temperature Dependence on Structure and Self-Diffusion of Water: A Molecular Dynamics Simulation Study using SPC/E Model. {\it Bull. Korean Chem. Soc.} {\bf 2013}, 34, 3800–3804.

\bibitem{Nafion_expt_D}Perrin, J.-C.; Lyonnard, S.; Volino, F. Quasielastic Neutron Scattering Study of Water Dynamics in Hydrated Nafion Membranes. {\it J. Phys. Chem. C} {\bf 2007}, 111, 3393–3404.

\end{thebibliography}
\end{document}